\documentclass[conference]{IEEEtran}
\ifCLASSOPTIONcompsoc
  \usepackage[nocompress]{cite}
\else
  \usepackage{cite}
\fi

\usepackage{amsmath,amssymb}
\usepackage{graphicx}
\usepackage{xurl}
\usepackage[hidelinks]{hyperref}
\usepackage{booktabs}
\usepackage{makecell}
\usepackage{tabularx}
\usepackage{float}
\usepackage{cleveref}
\usepackage{multirow}
\usepackage[table]{xcolor}
\usepackage{algorithmic}

\graphicspath{{figures/}}

\protected\def\mysys{\textsc{OVIG}}
\newcommand{\papertablefont}{\footnotesize}
\newcolumntype{L}{>{\raggedright\arraybackslash}X}
\newcolumntype{C}{>{\centering\arraybackslash}X}

\newcounter{paperalgorithm}
\renewcommand{\thepaperalgorithm}{\arabic{paperalgorithm}}
\crefname{paperalgorithm}{Algorithm}{Algorithms}
\Crefname{paperalgorithm}{Algorithm}{Algorithms}

\usepackage[HTML]{xcolor}
\definecolor{softGreen}{HTML}{D0E0D0}
\newcommand{\safe}[1]{{\setlength{\fboxsep}{0.5pt}\colorbox{softGreen}{#1}}}
\newcommand{\unsafe}[1]{#1}

\newcommand{\secref}[1]{Sec.~\ref{#1}}

\newtheorem{proposition}{Proposition}

\ifCLASSINFOpdf
\else
\fi

\begin{document}

\title{\mysys{}: Optimistic Verification of AI Training Integrity via Gradient Signals}

\author{\IEEEauthorblockN{
Hongxu Su\IEEEauthorrefmark{1},
Jianzhu Yao\IEEEauthorrefmark{2},
Huan Zhang\IEEEauthorrefmark{3},
Xuechao Wang\IEEEauthorrefmark{1}, and
Pramod Viswanath\IEEEauthorrefmark{2}}
\\[0.5pt]
\IEEEauthorblockA{\IEEEauthorrefmark{1}HKUST (GZ), Guangzhou, China\\
Email: hsu238@connect.hkust-gz.edu.cn, xuechaowang@hkust-gz.edu.cn}
\IEEEauthorblockA{\IEEEauthorrefmark{2}Princeton University, Princeton, NJ, USA\\
Email: jy0246@princeton.edu, pramodv@princeton.edu}
\IEEEauthorblockA{\IEEEauthorrefmark{3}University of Illinois Urbana-Champaign, Urbana, IL, USA\\
Email: huan@huan-zhang.com}
}


\maketitle

\begin{abstract}
The rapid growth of AI has increased the demand for domain-specific models. Post-training of open-source models offers a more economical way to meet this growing demand, but the cost of accelerator infrastructure often pushes organizations to outsource the process to third-party providers. An untrusted provider may deviate from the declared training procedure to save computation or inject malicious behavior. A potential solution is to audit the training by having an independent verifier replay the training and compare the results. This faces two key challenges: benign numerical drift from floating-point computation across heterogeneous accelerators is difficult to distinguish from malicious deviations, and the checkpoints and metadata  required for verification introduce substantial storage and transmission costs. Existing approaches either have security limitations or high deployment costs. We present \mysys{}, an optimistic verification framework that verifies training using an empirical boundary on gradient differences calibrated from honest heterogeneous replay. Gradient differences that exceed this boundary are treated as malicious deviations. \mysys{} further uses optimistic verification and partitions training into stride-$s$ intervals, retaining evidence only at interval endpoints to reduce storage and transmission costs. Across shortcut training attacks and adaptive targeted manipulation attacks, \mysys{} maintains \(0\%\) ASR on language, vision, and diffusion workloads. On Qwen3, increasing the stride from \(s=1\) to \(s=2000\) reduces off-chain storage and evidence transmission by \(1996\times\) while preserving \(0\%\) ASR and only adds \(14.3\%\) system computation overhead relative to unverified training. 
\end{abstract}

\IEEEpeerreviewmaketitle

\section{Introduction}
\label{sec:introduction}

As AI models are increasingly integrated into services and production systems, organizations across different domains are adapting them to domain-specific tasks. In areas such as healthcare, finance, and employment, training a foundation model from scratch is often impractical; instead, organizations typically post-train existing open-source models on their own data and requirements~\cite{warner2026omada,wong2025apoidea,handel2024indeed}. However, model training still requires costly accelerator infrastructure and substantial expertise to configure and operate the training stack. Organizations may therefore outsource post-training to third-party cloud or AI infrastructure providers~\cite{maslej2025artificial,mckinsey2025stateofaiagents,coreweave2025openai,mayoclinic2026microsoft,hampton2023cerebras}. An untrusted provider may deviate from the declared training procedure to reduce computation costs or manipulate training to introduce malicious behavior, such as a backdoor. Because the model owner has limited visibility into the actual execution, verifying whether the requested training was faithfully performed remains challenging.

A natural way to audit outsourced training is to evaluate the returned model on benchmarks. However, backdoor and Trojan attacks can be difficult to detect through such checks~\cite{liu2018trojaning, hong2022handcrafted, lv2023data}. Another way is through \emph{replay}: an independent verifier re-executes the declared training computation and checks whether the reproduced outcome is consistent with the provider's result. 
The central challenge is numerical variation from floating-point computation across heterogeneous hardware. Because floating-point arithmetic is non-associative under IEEE 754~\cite{IEEE8766229}, different accelerator reduction orders can produce slightly different results even during honest replay. Training verification must therefore distinguish benign numerical drift from malicious deviations.
Existing approaches mainly enforce deterministic execution~\cite{srivastava2024optimisticDeterministic}, inspect the trained model through watermarking or Zest-style tests~\cite{li2022verifying,zhang2024vtune,sander2024watermarking,ural2025securepol,jia2021entangled,maini2024llm,choi2023tools,jia2025backdoor}, compare model weights from provider and verifier using norm-based distances~\cite{jia2021proof}, or rely on cryptographic proofs~\cite{abbaszadeh2024kaizen} and hardware-assisted systems such as TEEs~\cite{duddu2024laminator,yuan2025cipherstealTEE,hornetz2026tdxrayTEE}. However, these approaches either incur substantial computation and deployment overhead or rely on signals and trust assumptions that leave exploitable security gaps.

To address this challenge, we present \mysys{}, an optimistic verification protocol that distinguishes benign numerical drift from malicious deviations by comparing gradients from provider's training and verifier's replay.
Our key observation is that honest heterogeneous replay produces small, bounded gradient differences, while malicious deviations can produce gradient differences that exceed this bound. This observation allows us to calibrate an empirical percentile boundary for verification.
In addition, to ensure that the provider's claimed training result is the same one examined during verification, \mysys{} anchors the claimed training trajectory on the blockchain before verification begins. This prevents the provider from modifying the training result during the audit to evade detection. 


\noindent\textbf{Robustness of the empirical boundary.} To construct the boundary, \mysys{} runs the same declared training procedure across heterogeneous devices, measures the resulting gradient differences, and forms a percentile profile from the envelope. We use gradients rather than weights because clipping, scaling, and other update operations can attenuate training deviations in weight space, and use percentile profiles rather than a scalar norm because scalar summaries can obscure deviations concentrated in a small subset of coordinates. Appendix~\ref{app:metric-rationale} further analyzes these design choices. 
We evaluate the boundary across language, vision, and diffusion models on ten GPUs under different training configurations. The calibrated boundary remains stable over time: honest gradient differences are typically larger early in training and decrease thereafter, while the boundary changes little as additional calibration steps are added. This allows \mysys{} to calibrate from a short early run and reuse the boundary for subsequent verification. Under fixed training configurations, the maximum observed honest false rejection rate (FRR) is below \(0.5\%\). We further evaluate boundary reuse across different LoRA layer ranges, datasets, input-length limits, and verifier GPUs, where the mean observed FRR is \(0.055\%\) and the maximum is \(1.1\%\). Details are reported in \secref{sec:raw-boundary-validation}. 

\noindent\textbf{Attack evaluation.} We evaluate \mysys{} against five classes of attacks. The first four represent common training deviations: microbatch dropping, stale updates, lower-precision training, and data-path perturbation. \mysys{} detects all evaluated instances of these attacks, yielding \(0\%\) ASR. The last one is an adaptive attack based on PGD, which is a representative and strong white-box attack~\cite{madry2017towards,croce2021mind}. It introduces a chosen gradient direction to manipulate selected model behavior, as detailed in \secref{sec:target-perturbation-attack}. The attack can alter targeted predictions while leaving aggregate benchmark performance nearly unchanged (\secref{sec:benchmark-level-evidence}). We compare \mysys{} with PoTD~\cite{choi2023tools}, RTTD~\cite{jia2025backdoor}, and PoL~\cite{jia2021proof} under this adaptive attack. \mysys{} maintains \(0\%\) ASR across all evaluated settings, whereas each baseline is bypassed in at least one setting. Detailed results are reported in \secref{sec:baseline-comparison}.

\noindent\textbf{Verification cost.}
Verifier replay introduces additional computation cost. \mysys{} adopts optimistic verification, which accepts the provider's training result by default and randomly checks only portions of the training process. This avoids replaying the entire trajectory and substantially reduces verification cost. Security is supported by deposit-and-slashing mechanisms widely used in accountable blockchain protocols such as Casper~\cite{buterin2017casper}. In \mysys{}, verification is performed by a committee assuming that a majority of its members are honest. Committee construction is orthogonal to our design and can rely on existing protocols~\cite{gilad2017algorand,david2022gearbox}. Because boundary calibration is a one-time setup cost performed before provider execution, it is excluded from the reported system computation overhead. Under this setting, \mysys{} adds only \(14.3\%\) system computation overhead relative to unverified training, as detailed in
\secref{sec:consumption-tradeoff}. 

\noindent\textbf{Storage and transmission cost.}
Another source of overhead is the \emph{evidence} that the provider must store and transmit to the verifier, mainly including checkpoints used for verification and metadata required for replay. To reduce this cost, \mysys{} partitions the training trajectory into stride-$s$ intervals and retains only evidence of endpoints. A larger $s$ reduces storage and transmission, but also increases the replay length and may require a looser boundary to avoid honest rejection. This creates a tradeoff between storage and transmission cost and security. 
In \secref{sec:baseline-comparison}, PoL admits successful attacks at $s=1$, whereas \mysys{} maintains $0\%$ ASR even at $s=2000$. At $s=2000$, \mysys{} reduces storage and evidence transmission by $1996\times$ relative to $s=1$ on Qwen3. Compared with deterministic replay~\cite{srivastava2024optimisticDeterministic}, \mysys{} reduces additional system computation overhead by $4.75\times$ and storage and evidence transmission by $1148.94\times$.

Our main contributions are as follows:
\begin{itemize}
    \item We propose an empirical boundary calibrated from gradient differences to distinguish benign numerical drift from malicious deviations during training.

    \item We design \mysys{}, a deployable training verification protocol that combines the empirical boundary with optimistic verification and sparse storage evidence.

    \item We evaluate \mysys{} across heterogeneous GPUs and language, vision, and diffusion workloads. The results show that \mysys{} detects all evaluated attacks while providing practical verification and storage efficiency.
\end{itemize}




\section{Related Work}
\label{sec:related-work}

\noindent\textbf{Behavioral and data usage evidence.}
Several lightweight approaches verify outsourced training through final model behavior or data-use signals. Benchmark-level checks infer whether the returned model has plausible task utility~\cite{li2022verifying}. Watermarking, radioactive data, entangled data, and backdoor-driven verifiable fine-tuning embed evidence into the data or training procedure that can later be queried from the model~\cite{sander2024watermarking,jia2021entangled,zhang2024vtune,maini2024llm,ural2025securepol}. Memorization-based methods such as PoTD use the tendency of models to overfit recent examples as evidence of data exposure~\cite{choi2023tools}. Zest tests similarly reason from final model behavior~\cite{jia2025backdoor}.

These methods are useful for accountability, ownership, and data usage verification. They do not by themselves establish that the provider followed the declared training process. A provider may use plausible data, preserve watermark triggers, and retain aggregate performance while skipping work, altering numerical paths, or applying localized target-directed perturbations. Such deviations may remain undetected under checks that observe only final behavior or data-exposure imprints.

\noindent\textbf{Training process replay.} 
Auditing the training process by replaying provides more direct evidence of training execution. Proof-of-Learning records training metadata, batch indices, and intermediate states so that a verifier can replay selected updates~\cite{jia2021proof}. Related systems use redundancy, provenance, or fraud proof interaction to reduce the amount of computation that must be checked~\cite{jia2025backdoor,chang2025towards,teutsch2024scalable,conway2024opml}. These approaches move beyond final model inspection, but many replay checks summarize an interval through scalar endpoint distances. Such summaries can be vulnerable to attacks that exploit permissive distance margins, especially as replay intervals grow~\cite{fang2023proof}.

Exact replay also faces a systems problem. IEEE 754 arithmetic, reduction order, fused kernels, scheduling, and mixed precision can differ across heterogeneous accelerators. Bitwise deterministic replay is therefore difficult to deploy in optimized training environments. Prior work can improve determinism through controlled kernels, rounding logs, or constrained numerical paths~\cite{zhuang2022randomness,srivastava2024optimisticDeterministic,chen2022towards,xu2022checkpointing}. Those techniques strengthen reproducibility, but they can also increase cost or change the natural training environment. \mysys{} instead checks the gradient differences produced by heterogeneous replay while explicitly tolerating honest heterogeneous drift.

\noindent\textbf{Cryptographic and hardware assisted verification.} 
Cryptographic proof systems can provide strong arithmetic-level guarantees for outsourced computation. Zero knowledge proof and proof-of-training systems are therefore attractive for high-assurance settings~\cite{abbaszadeh2024kaizen,sun2024zkdl}. For routine large-model training, however, current proof systems remain costly and difficult to deploy at scale~\cite{garg2023experimenting,tan2025founding,liang2025sok}.

Trusted execution environments offer a different tradeoff by binding execution metadata and artifacts to attested hardware~\cite{duddu2024laminator}. Their efficiency is appealing, but their guarantees depend on hardware isolation, attestation, and side-channel assumptions. Recent work shows that optimized AI workloads in trusted hardware can still face microarchitectural or host-level risks~\cite{yuan2025cipherstealTEE,hornetz2026tdxrayTEE}, making hardware attestation an important but incomplete trust boundary.

\noindent\textbf{Positioning.} 
\mysys{} verifies training using gradient differences, tolerates benign numerical variation with an empirical boundary, and reduces verification cost through optimistic auditing and sparse endpoint evidence.

\section{Problem Formulation and Threat Model}
\label{sec:threat-model}

\subsection{System Setting}
\label{sec:system-setting-parties}

We consider outsourced post-training with three parties: a model owner
\(\mathcal U\), an untrusted provider \(\mathcal P\), and an audit committee
\(\mathcal K\) with an odd number of members \(n_{\mathcal K}\). Before
provider execution, the model owner publishes a public task
\begin{equation}
\mathcal T =
\left(
\mathcal M_0, D, \Pi, \Omega, N, s, m, \mathcal B^{(s)}
\right),
\label{eq:threat-public-task}
\end{equation}
where \(\mathcal M_0\) is the initial checkpoint, \(D\) is the training dataset, \(\Pi\) is the training policy, \(\Omega\) is the replay metadata, \(N\) is the number of training steps, \(s\) is the replay stride, \(m\) denotes the trainable parameter set, and \(\mathcal B^{(s)}\) is the deployed stride-specific boundary. The trainable parameters \(m\) may consist of model parameters or parameter-efficient adapters such as LoRA. Together, \((D,\Pi,\Omega)\) determine the declared batch schedule, optimizer, update rule, randomness, numerical configuration, and other information required to replay the training computation.

\noindent\textbf{Verification scope.} For a task $\mathcal T$, we define the set of honest training trajectories as 
\begin{equation*} 
\mathcal H(\mathcal T)= \left\{ \tau : \tau \text{ follows } \mathcal T \text{on supported hardware} \right\}. 
\end{equation*} 
Thus, honest executions may produce numerically different trajectories across heterogeneous hardware, but all remain in $\mathcal H(\mathcal T)$ as long as they follow the declared training procedure. \mysys{} targets trajectories $\tau \notin \mathcal H(\mathcal T)$ whose gradient differences exceed the calibrated boundary during replay, termed \emph{gradient-visible deviations}.

The public task $\mathcal T$ fixes the declared training procedure. After training, the provider commits the final values $m_f$ of the trainable parameters together with the evidence required for replay verification. For each stride interval $i$, we denote the retained evidence by
\begin{equation}
\mathcal E_i =
\left(
m_{a_i},\, m_{b_i},\, \Omega_i^{\mathrm{rep}},\, Z_{a_i}
\right),
\end{equation}
where $m_{a_i}$ and $m_{b_i}$ are the trainable parameters at the interval endpoints, $\Omega_i^{\mathrm{rep}}$ contains the interval-specific metadata required for replay, and $Z_{a_i}$ denotes any additional training state needed to resume execution from $a_i$, such as optimizer state for AdamW; $Z_{a_i}=\varnothing$ when no such state is required. Parameters outside $m$ are inherited from $\mathcal M_0$ and remain fixed according to the declared task. The provider commits $m_f$ and $\{\mathcal E_i\}$ before any audit-related randomness is revealed; subsequent randomness only determines the audited intervals and, when enabled, the sampled gradient coordinates. \mysys{} verifies the declared training computation using this committed evidence, while attacks that directly target the optimizer state itself are outside the scope of this paper.

A stride \(s\) induces \(K_s\) audit intervals, defined in \secref{sec:stride-based-replay}. We use \(\mathsf{VerifyInterval}(i)\) for the interval-level audit predicate defined in \secref{sec:protocol-opening-verification}.

\subsection{Adversary Goals and Capabilities} \label{sec:adversary-goals} The adversary is an untrusted provider \(\mathcal P^\star\). Before committing its evidence, \(\mathcal P^\star\) knows the task specification \(\mathcal T\) and the verification rules. It controls the training environment such as hardware, kernels, dataset, and off-chain evidence generation. Before commitment, \(\mathcal P^\star\) may execute any training path that deviates from the declared computation. It may also commit endpoint evidence that is inconsistent with the execution it actually performed. After commitment, it cannot change committed objects or predict the audit randomness, except by breaking assumptions in A1 and A2.

Let \(\Delta_{\mathcal P}\) denote the set of provider executions that deviate from the compliant execution set \(\mathcal H(\mathcal T)\): 
\begin{equation} 
    \Delta_{\mathcal P} = \left\{ 
        \tau^\star : 
        \tau^\star \notin \mathcal H(\mathcal T) 
    \right\}. 
    \label{eq:provider-deviation-set} 
\end{equation}

\mysys{} does not aim to detect every deviation in \(\Delta_{\mathcal P}\). Instead, it targets the subset of \emph{gradient-visible deviations}, 
\begin{equation} 
    \Delta_{\mathcal P}^{\mathrm{grad}} \subseteq \Delta_{\mathcal P}. 
    \label{eq:gradient-visible-deviation-set} 
\end{equation} 
Here, \(\Delta_{\mathcal P}^{\mathrm{grad}}\) denotes deviations that produce a detectable difference between the replayed and provider-claimed gradients of the trainable parameter subset \(m\). Such deviations are rejected when this difference exceeds the deployed boundary \(\mathcal B^{(s)}\).

For deviations covered by \mysys{}, the adversary succeeds if it commits a deviating execution \(\tau^\star \in \Delta_{\mathcal P}^{\mathrm{grad}}\) that is nevertheless accepted by the optimistic audit. For an audit set \(\mathcal Q\), this requires 
\begin{equation} 
    \tau^\star \in \Delta_{\mathcal P}^{\mathrm{grad}} \;
    \wedge\; \forall i \in \mathcal Q:\; 
    \mathsf{VerifyInterval}(i)=\textsc{accept}. 
    \label{eq:adversary-success} 
\end{equation}

Our security evaluation considers two adversarial settings. First, we evaluate whether \mysys{} detects common noncompliant executions, including skipped batches, stale gradients, shortened computation, undeclared precision changes, and data-path tampering. These experiments test whether such deviations become gradient-visible and are rejected by the deployed boundary. Second, we consider an adaptive adversary that knows the verification rule and explicitly constrains its gradient perturbation to remain within $\mathcal B^{(s)}$. Subject to passing verification, the adversary maximizes a target-directed perturbation and attempts to alter selected model behavior. This adaptive setting therefore tests whether a deviation that evades the boundary check can still manipulate the model behavior. 


\subsection{Trust Assumptions}
\label{sec:trust-assumptions}

\noindent\textbf{A1: Binding commitments.} 
We assume that Merkle commitments are binding and that commitments recorded on the blockchain cannot be altered. Therefore, once the provider posts its commitments, it cannot change the committed evidence without detection.

\noindent\textbf{A2: Unpredictable audit selection.}
Intervals are randomly selected for verification after the provider has completed training and posted its commitments. The provider cannot predict which intervals will be verified. 

\noindent\textbf{A3: Honest committee majority.}
We assume that a majority of committee members are honest and follow the prescribed verification procedure. Attacks that corrupt a majority of the committee are outside the scope of this paper.

\noindent\textbf{A4: Honest boundary calibration.}
The deployed boundary \(\mathcal B^{(s)}\) is honestly calibrated and committed on-chain before provider execution. The model owner may delegate calibration to a third party. The calibration process will be verified by committee under A3. 

\noindent\textbf{A5: Evidence availability and disclosure.}
The provider must keep retained endpoint evidence and any data needed for opened replay available off chain. This paper does not provide data privacy during disputes.

\noindent\textbf{A6: Rational participants.}
We assume participants are economically rational and choose strategies that maximize expected payoff under the settlement rule.

\subsection{Security Goals}
\label{sec:security-goals}

\noindent\textbf{G1: Acceptance for Honest Providers.}
An honest interval may rarely be rejected because of benign numerical drift under heterogeneous replay. For an audit that samples \(q\) intervals, we denote the resulting audit-level FRR by
\(\eta_{\mathrm{fr}}\):
\begin{equation}
\Pr
\left[
\forall i\in\mathcal Q:
\mathsf{VerifyInterval}(i)=\mathsf{ACCEPT}
\right]
=
1-\eta_{\mathrm{fr}}.
\label{eq:audit-frr}
\end{equation}

For an honest training trace containing \(K_s\) intervals, if
\(R_{\mathrm{fr}}\) intervals are rejected under the deployed boundary,
uniform sampling of \(q\) intervals without replacement gives the
empirical audit-level FRR
\begin{equation}
\widehat{\eta}_{\mathrm{fr}}
=
1-
\frac{
\binom{K_s-R_{\mathrm{fr}}}{q}
}{
\binom{K_s}{q}
}.
\label{eq:empirical-audit-frr}
\end{equation}

\noindent\textbf{G2: Detecting Attacks on Opened Intervals.} For each opened interval \(i\), the committee applies \(\mathsf{VerifyInterval}(i)\) using the percentile profiles and boundary check defined in Algorithm~\ref{alg:stride-audit}. The goal is to reject gradient-visible deviations whose replayed gradient differences exceed \(\mathcal B^{(s)}\). For deviations that remain within the boundary, we evaluate whether the admissible perturbation is sufficient to manipulate model behaviors. 

\noindent\textbf{G3: Incentive Compatibility of Optimistic Audit.} Under the rational-participant assumption, the optimistic audit should make honest execution the provider's payoff-maximizing strategy. The goal is not to audit every interval, but to make any detectable deviation economically unattractive before the provider commits its evidence. For a sampled audit set \(\mathcal Q\), the reward, deposit, slashing amount, and audit probability should be chosen so that a malicious provider has lower expected payoff than an honest provider. This requires the detection probability of the deviation to exceed the FRR, and the deposit to cover the provider's avoided training cost, committee compensation, and private benefit from successful deviation. The settlement condition is formalized in \secref{sec:commitment-settlement}.

\section{\mysys{} Design}
\label{sec:protocol-design}

\subsection{Overview}
\label{sec:design-overview}
\mysys{} involves three roles: a \emph{model owner}, a \emph{provider}, and a \emph{committee}. The model owner specifies the training task and empirical boundary. The provider performs the training and commits the training trajectory. The committee verifies the provider's execution by replaying the declared training procedure and applying the empirical boundary check.

\begin{figure*}[t]
\centering
\includegraphics[width=0.88\linewidth]{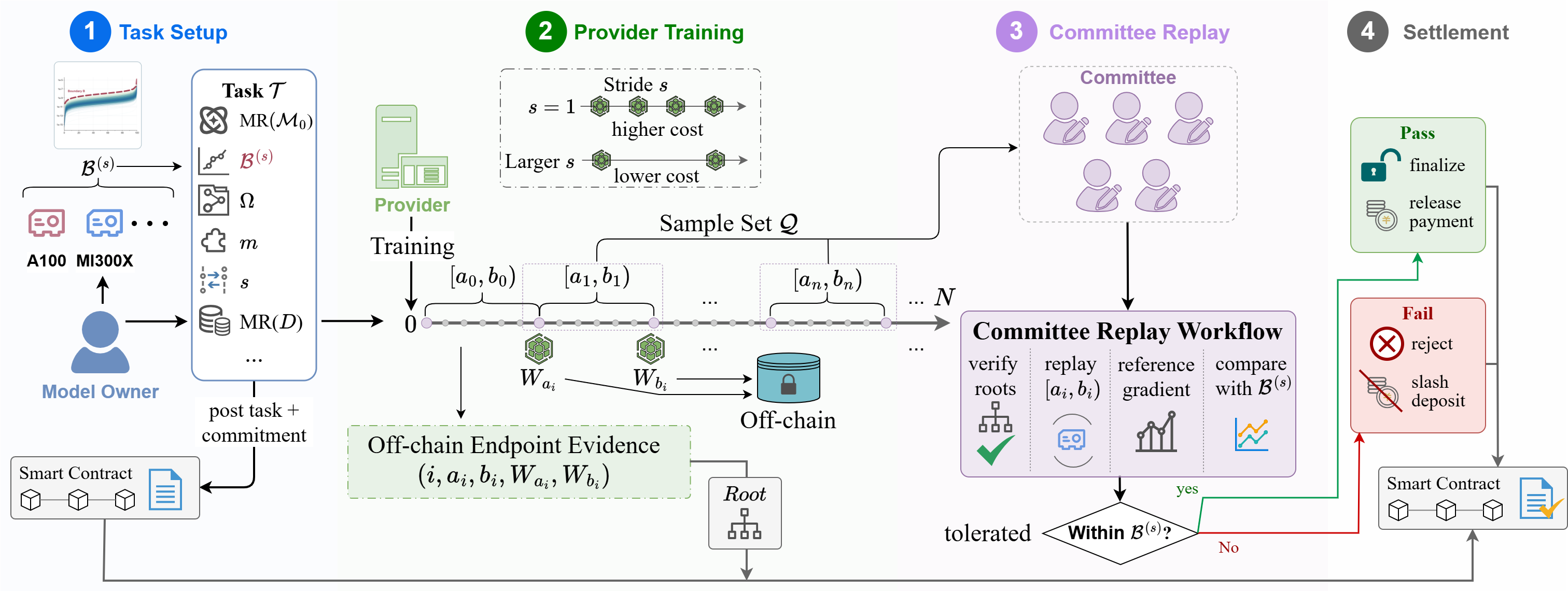}
\caption{\mysys{} protocol workflow, following the four-stage design in \secref{sec:design-overview}.}
\label{fig:protocol-workflow}
\end{figure*}

The design has four steps. Figure~\ref{fig:protocol-workflow} summarizes the workflow. 
\begin{enumerate}
\item \textbf{Task setup and public commitment.} The owner publishes a fully specified task: the initial model, dataset, training policy, replay metadata, stride, and boundary.

\item \textbf{Provider training and commitment.} The provider trains according to this public specification and retains the replay evidence at stride-aligned endpoints, including the trainable parameters and any optimizer state required for replay. After training, it posts the training process commitment and final model checkpoint commitment.

\item \textbf{Random audit and committee replay.} After the provider commits the training trajectory, a random subset of the training process is selected for audit. The committee replays the selected computation and checks its consistency with the committed trajectory using the empirical boundary.

\item \textbf{Mechanism and settlement.} Finally, settlement follows the optimistic result: all sampled intervals must pass for the provider to be paid.
\end{enumerate}

\subsection{Task Setup and Public Commitment}
\label{sec:task-setup-commitment}

We use one generic notation for Merkle commitments throughout the protocol. For
an ordered object \(X=(X_0,\ldots,X_{n-1})\), let
\begin{equation}
\mathsf{MR}(X)
=
\mathsf{MerkleRoot}
\big(
H(0\,\|\,X_0),\ldots,H(n-1\,\|\,X_{n-1})
\big).
\label{eq:mr-definition}
\end{equation}
Here \(\mathsf{MR}(X)\) is the Merkle-tree root of a canonical serialization of \(X\). For large tensors or datasets, \(X_i\) denotes a chunk or record. A Merkle opening later proves that an opened element is part of the committed ordered object.

\noindent\textbf{On-chain and off-chain data.}
\mysys{} stores only lightweight protocol data directly on the blockchain, which we refer to as \emph{on-chain} data. Large objects, such as model checkpoints, optimizer states, and training data, remain \emph{off-chain}; only their Merkle roots are committed on chain. 

The model owner publishes the task before provider execution. The task contains
the initial model \(\mathcal M_0\), the training dataset \(D\), a training
policy \(\Pi\), replay metadata \(\Omega\), the number of training steps \(N\),
the replay stride \(s\), the trainable parameter subset \(m\), and the deployed
boundary \(\mathcal B^{(s)}\):
\begin{equation}
\mathcal T
=
\left(
\mathcal M_0,
D,
\Pi,
\Omega,
N,
s,
m,
\mathcal B^{(s)}
\right).
\label{eq:public-task}
\end{equation}
The policy \(\Pi\) specifies the training rule, including the optimizer, loss
function, batch size, stopping rule, preprocessing, and aggregation rule. The
metadata \(\Omega\) specifies replay-critical values such as random seeds, data
ordering, learning-rate schedule, precision policy, and any step-local
configuration needed to reconstruct the same batch and update rule.

The on-chain task publication contains
\begin{equation}
\left(
\mathsf{MR}(\mathcal M_0),
\mathsf{MR}(D),
\Pi,
\Omega,
N,
s,
m,
\mathcal B^{(s)}
\right).
\label{eq:onchain-task}
\end{equation}
The initial model and dataset are large, so only their Merkle roots are placed on chain. The policy, metadata, stride, trainable parameter subset, and boundary are small and are published directly. The corresponding off-chain objects are public and authenticated by the posted roots. Thus, before training starts, both the provider and the committee can check the task, reconstruct the batch schedule, and prepare replay.

This setup fixes the claimed training path. For every step \(t\), the batch index, random seed, optimizer configuration, learning rate, and other step-local values are determined by \((D,\Pi,\Omega,t)\). The committee does not introduce any training randomness. Its later audit randomness only chooses which already-fixed intervals are opened.

The boundary published by the owner is
\begin{equation}
\mathcal B^{(s)}
=
\left\{
B_s^{\mathrm{abs}},
B_s^{\mathrm{rel}}
\right\},
\label{eq:boundary-object}
\end{equation}
where \(B_s^{\mathrm{abs}}\) and \(B_s^{\mathrm{rel}}\) are percentile profiles
for absolute and relative gradient differences.

\subsection{Stride Aligned Replay}
\label{sec:stride-based-replay}

The stride \(s\) partitions the \(N\)-step training run into
\(\lceil N/s\rceil\) replay intervals. Let
\begin{equation}
K_s=\lceil N/s\rceil .
\label{eq:num-stride-intervals}
\end{equation}
For \(i=0,1,\ldots,K_s-1\), the start and end indices of the \(i\)-th interval
are
\begin{equation}
a_i=i\cdot s,
\qquad
b_i=\min\{(i+1)\cdot s,N\}.
\label{eq:stride-intervals}
\end{equation}

For replay, we associate each stride-aligned endpoint with a training state
\begin{equation}
S_t=\left(W_t^{(m)},Z_t\right),
\label{eq:training-state}
\end{equation}
where \(W_t^{(m)}\) denotes the trainable parameters and \(Z_t\) contains
any optimizer state required to resume training. For stateless optimizers,
\(Z_t=\varnothing\).

The provider's claimed trajectory is represented by a single ordered sequence
of committed endpoint states. Let
\begin{equation}
t_i=\min\{i\cdot s,N\},
\qquad
i=0,1,\ldots,K_s,
\label{eq:stride-endpoint-indices}
\end{equation}
such that \(a_i=t_i\) and \(b_i=t_{i+1}\). The committed endpoint sequence is
\begin{equation}
\Gamma^c
=
\left(
S_{t_0}^c,
S_{t_1}^c,
\ldots,
S_{t_{K_s}}^c
\right),
\qquad
S_{t_i}^c
=
\left(
W_{t_i}^{c,(m)},
Z_{t_i}^c
\right).
\label{eq:committed-endpoint-sequence}
\end{equation}
Accordingly, interval \(i\) is defined by the two adjacent committed states
\begin{equation}
\left(S_{a_i}^c,S_{b_i}^c\right)
=
\left(S_{t_i}^c,S_{t_{i+1}}^c\right).
\label{eq:linked-replay-interval}
\end{equation}
Thus, interval endpoints are not independently supplied: the endpoint of
interval \(i\) is the same committed state as the starting point of interval
\(i+1\).

The sequence is anchored to the declared initial checkpoint and the returned
final checkpoint:
\begin{equation}
W_{t_0}^{c,(m)}
=
\operatorname{Extract}_{m}(\mathcal M_0),
\qquad
W_{t_{K_s}}^{c,(m)}
=
\operatorname{Extract}_{m}(\mathcal M_N^c).
\label{eq:trajectory-endpoint-binding}
\end{equation}
Since parameters outside \(m\) are fixed by the public task, the returned
checkpoint must additionally satisfy
\begin{equation}
\operatorname{Extract}_{\bar m}(\mathcal M_N^c)
=
\operatorname{Extract}_{\bar m}(\mathcal M_0).
\label{eq:frozen-parameter-binding}
\end{equation}

The provider retains these stride-aligned endpoint states for later opening.
Stride is therefore a deployment parameter that trades verification
granularity for evidence cost; Sec.~\ref{sec:consumption-tradeoff} provides
practical guidance for selecting \(s\).

For readability, we omit the superscript \((m)\) below when the trainable parameter subset is clear.

For an opened interval \([a_i,b_i)\), the provider reveals authenticated
openings of the adjacent committed states
\(\widehat S_{a_i}\) and \(\widehat S_{b_i}\) from \(\Gamma^c\).
Replay reruns the declared training updates from the opened start state on a
committee device while holding all non-device inputs fixed. We write
\begin{equation}
\mathsf{Replay}
\left(
h,
\widehat S_{a_i},
a_i,
b_i;
\mathcal T
\right)
\rightarrow
\widetilde S_{b_i},
\label{eq:replay-function}
\end{equation}
where \(h\) is the committee accelerator,
\(\widehat S_{a_i}
=
(\widehat W_{a_i}^{(m)},\widehat Z_{a_i})\)
is the authenticated committed start state, and \(\mathcal T\) fixes the
optimizer, loss, batch schedule, random seeds, and all other replay inputs.
The output
\(\widetilde S_{b_i}
=
(\widetilde W_{b_i}^{(m)},\widetilde Z_{b_i})\)
is the endpoint state produced by the committee's replay.

The committee then computes endpoint gradients at both the replayed endpoint
\(\widetilde W_{b_i}\) and the authenticated committed endpoint
\(\widehat W_{b_i}\):
\begin{equation}
\begin{aligned}
G'_{b_i}
&=
\nabla_W^{(m)}
\mathcal L_{\mathcal T}
\left(
\widetilde W_{b_i},
I_{b_i};
\Pi,\Omega
\right),
\\
G^*_{b_i}
&=
\nabla_W^{(m)}
\mathcal L_{\mathcal T}
\left(
\widehat W_{b_i},
I_{b_i};
\Pi,\Omega
\right).
\end{aligned}
\label{eq:replay-endpoint-gradient}
\end{equation}
Here, \(\nabla_W^{(m)}\) denotes the gradient operator with respect to the trainable parameters, \(\mathcal L_{\mathcal T}\) is the loss function specified in \(\Pi\), and \(I_{b_i}\) is the batch used for the endpoint check. If \(b_i=N\), the endpoint-check batch is derived by the same public rule specified in \(\Omega\).

The committee then flattens both gradients:
\begin{equation}
x'_i=\operatorname{Flatten}(G'_{b_i}),
\qquad
x^*_i=\operatorname{Flatten}(G^*_{b_i}).
\label{eq:flatten-gradients}
\end{equation}
Their coordinate-wise difference is
\begin{equation}
\Delta_{b_i}
=
|x'_i-x^*_i|.
\label{eq:gradient-difference}
\end{equation}
An honest provider and an honest committee may still produce small differences
because replay devices, kernels, reduction orders, and floating-point paths
can differ. The empirical boundary defined below determines which differences
are acceptable.

\subsection{Empirical Boundary Calibration}
\label{sec:protocol-boundary-calibration}

The empirical boundary is calibrated before provider execution. The model owner may delegate calibration to the audit committee or an independent calibration service with access to heterogeneous accelerators. The calibration process is also public for verification by committee. Calibration is a one-time setup cost and the resulting boundary can be reused for subsequent audits. 

For a target stride \(s\), calibration first runs honest training for a small
number of intervals, for example five stride intervals (\(5s\) training steps). This produces honest interval states $(S_{a_i},S_{b_i})$. Let
\[
\mathcal H=\{h_1,\ldots,h_M\}
\]
be the set of heterogeneous accelerators used for calibration. For each
calibration interval \(i\) and device \(h\in\mathcal H\), the calibrator runs
\(\mathsf{Replay}\) from \(S_{a_i}\) to \(b_i\), computes the replayed endpoint gradient, loads \(W_{b_i}\), and computes the corresponding endpoint gradient on the same batch. Let
\[
x'_{i,h}
=
\operatorname{Flatten}(G'_{b_i,h}),
\qquad
x^*_{i,h}
=
\operatorname{Flatten}(G^*_{b_i,h}).
\]

Let \(d\) denote the number of gradient coordinates in the trainable parameter set. For coordinate \(j\in[1,d]\), the maximum honest absolute deviation is
\begin{equation}
\overline{\Delta}^{\mathrm{abs}}_j
=
\max_{i,h}
\left|
x'_{i,h,j}
-
x^*_{i,h,j}
\right|.
\label{eq:max-absolute-error}
\end{equation}
The corresponding relative deviation is normalized by the largest observed
honest gradient magnitude at that coordinate:
\begin{equation}
\overline{\Delta}^{\rm rel}_j
=
\max_{i,h}
\frac{|x'_{i,h,j}-x^*_{i,h,j}|}
{\max\{|x'_{i,h,j}|,|x^*_{i,h,j}|\}+\epsilon}.
\label{eq:relative-error}
\end{equation}
where \(\epsilon>0\) is a small constant avoiding division by zero.

The boundary is represented as percentile profiles.
Let
\begin{equation}
\Lambda=\{1,2,5,10,15,\ldots,95,98,100\}\subset[0,100]
\label{eq:percentile-grid}
\end{equation}
be the percentile grid, and let \(Q_p(\cdot)\) denote the empirical
\(p\)-th percentile where \(p\in\Lambda\). The raw absolute and relative profiles are
\begin{equation}
\begin{aligned}
\widetilde B_s^{\mathrm{abs}}(p)
&=
Q_p
\left(
\{\overline{\Delta}^{\mathrm{abs}}_j\}_{j=1}^d
\right),
\\
\widetilde B_s^{\mathrm{rel}}(p)
&=
Q_p
\left(
\{\overline{\Delta}^{\mathrm{rel}}_j\}_{j=1}^d
\right). 
\end{aligned}
\label{eq:raw-boundary}
\end{equation}
Deployment inflates the raw boundary by a safety factor
\(\alpha_{\mathcal B}>1\):
\begin{equation}
B_s^{\mathrm{abs}}(p)
=
\alpha_{\mathcal B}
\cdot
\widetilde B_s^{\mathrm{abs}}(p),
\qquad
B_s^{\mathrm{rel}}(p)
=
\alpha_{\mathcal B}
\cdot
\widetilde B_s^{\mathrm{rel}}(p).
\label{eq:inflated-boundary}
\end{equation}
The committed boundary is then
\[
\mathcal B^{(s)}
=
\{B_s^{\mathrm{abs}},B_s^{\mathrm{rel}}\}.
\]

For a large trainable parameter subset \(m\), computing all \(d\) coordinates may be unnecessary. The calibration and verification procedures may instead compute the empirical percentiles on a uniformly sampled coordinate subset. The sample size is fixed. During verification, the coordinate sampling seed is derived from the provider's posted commitments, analogously to $\rho_{\mathrm{aud}}$. This prevents the provider from predicting which gradient coordinates will be checked during verification. We analyze the sampling error in Appendix~\ref{app:sampling-proof}.

\subsection{Provider Training and Commitment}
\label{sec:provider-training-commitment}

After accepting a task, the provider trains for \(N\) steps using the public task \(\mathcal T\). Since \(\mathcal T\) fixes the training path, both the provider and the committee can reconstruct the batch schedule and replay procedure for any interval before training starts. The provider only needs to execute the declared path.

During training, the provider retains the stride-aligned endpoint states in the ordered sequence \(\Gamma^c\). After training, it posts
\begin{equation}
\mathsf{MR}(\widehat{\mathcal M}_N),
\qquad
\mathsf{MR}\left(
\{(i,t_i,S_{t_i}^c)\}_{i=0}^{K_s}
\right).
\label{eq:provider-commitments}
\end{equation}
The full parameter tensors and optimizer states remain off chain. During an audit, Merkle openings authenticate that the revealed interval endpoints are exactly those committed after training. This commitment prevents the provider from changing the opened endpoints after seeing the sampled intervals. Before interval sampling, the protocol additionally verifies the two trajectory anchors in Eq.~\eqref{eq:trajectory-endpoint-binding} and Eq.~\eqref{eq:frozen-parameter-binding}.

\subsection{Random Audit and Committee Replay}
\label{sec:protocol-opening-verification}

After the provider posts the commitment to the final model and endpoint evidence, the protocol samples a subset of stride intervals for audit. Let \(\phi\in(0,1]\) be the audit fraction. Since stride \(s\) induces \(K_s=\lceil N/s\rceil\) intervals, the number of sampled intervals is
\begin{equation}
q=\lceil \phi K_s\rceil .
\label{eq:audit-budget}
\end{equation}

The audit set is sampled only after the provider has fixed its committed evidence. Let \(R_{\mathrm{state}}\) denote the Merkle root of the committed endpoint states in \cref{eq:provider-commitments}. The audit seed is derived as 
\begin{equation}
\rho_{\mathrm{aud}}
=
H\left(
\textsf{audit}
\,\|\,\mathsf{MR}(\widehat{\mathcal M}_N)
\,\|\,R_{\mathrm{state}}
\,\|\,\rho_{\mathrm{post}}
\right),
\label{eq:audit-randomness}
\end{equation}
where \(\rho_{\mathrm{post}}\) may be a block hash or public randomness beacon
available only after the provider commitments are posted. The sampled interval
set is
\begin{equation}
\mathcal Q
\leftarrow
\mathsf{SampleIntervals}(K_s,q;\rho_{\mathrm{aud}}).
\label{eq:audit-set}
\end{equation}
This randomness only chooses which intervals are opened; it does not affect the replay configuration.

For every sampled interval \(i\in\mathcal Q\), the provider opens the committed replay evidence $\mathcal E_i$ with the corresponding Merkle authentication paths. We write \(\mathsf{VerifyInterval}(i)\) for the committee procedure in Algorithm~\ref{alg:stride-audit}. Each committee member independently executes \(\mathsf{VerifyInterval}\) for every
\(i\in\mathcal Q\). The audit accepts the provider's claim only if a majority of committee members vote to accept; any rejected sampled interval triggers the settlement rule in \secref{sec:commitment-settlement}.

\begin{figure}[t]
\centering
\footnotesize
\refstepcounter{paperalgorithm}\label{alg:stride-audit}
\begin{minipage}{0.98\linewidth}
\hrule
\vspace{0.35em}
{\raggedright\noindent\textbf{Algorithm~\thepaperalgorithm: \(\mathsf{VerifyInterval}(i)\), Committee Audit of Interval $i$}\par}
\vspace{0.25em}
\begin{algorithmic}[1]
\REQUIRE Public task \(\mathcal T\), sampled interval \(i\), opened replay evidence $\widehat{\mathcal E}_i$, Merkle paths, committee device \(h\).
\ENSURE \textsc{accept} or \textsc{reject}.

\STATE Check that \(a_i=i\cdot s\) and \(b_i=\min\{(i+1)s,N\}\).

\STATE Authenticate \(\widehat S_{a_i}\) and \(\widehat S_{b_i}\) against
\(R_{\mathrm{state}}\) at positions \(i\) and \(i+1\), respectively.

\STATE Authenticate \(S_{0}^{c}\) and \(S_{N}^{c}\) against
\(R_{\mathrm{state}}\), and check
\[
W_{0}^{c,(m)}=\operatorname{Extract}_{m}(\mathcal M_0),
\qquad
W_{N}^{c,(m)}=\operatorname{Extract}_{m}(\mathcal M_N^c),
\]
\[
\operatorname{Extract}_{\bar m}(\mathcal M_N^c)
=
\operatorname{Extract}_{\bar m}(\mathcal M_0).
\]

\STATE Authenticate the required model and data chunks against
\(\mathsf{MR}(\mathcal M_0)\) and \(\mathsf{MR}(D)\).

\STATE Reconstruct the interval batch schedule from \((D,\Pi,\Omega)\).

\STATE Run
\[
\mathsf{Replay}(h,\widehat S_{a_i},a_i,b_i;\mathcal T)
\]
to obtain \(\widetilde S_{b_i}\).

\STATE If \(Z_{b_i}\neq\varnothing\), check
\[
\mathsf{Consistent}_{Z}
\left(
\widetilde Z_{b_i},
\widehat Z_{b_i}
\right)=1.
\]

\STATE Compute endpoint gradients \(G'_{b_i}\) at \(\widetilde W_{b_i}\) and \(G^*_{b_i}\) at \(\widehat W_{b_i}\) on the same endpoint-check batch.

\STATE Let \(\mathcal C\) be all gradient coordinates, or the sampled coordinate subset if coordinate sampling is enabled.

\STATE Set \(x'=\operatorname{Flatten}(G'_{b_i})_{\mathcal C}\) and \(x^*=\operatorname{Flatten}(G^*_{b_i})_{\mathcal C}\).

\STATE For every \(p\in\Lambda\), compute
\[
P_i^{\mathrm{abs}}(p)
=
Q_p\left(
\{|x'_j-x^*_j|\}_{j\in\mathcal C}
\right).
\]

\STATE For every \(p\in\Lambda\), compute
\[
P_i^{\mathrm{rel}}(p)
=
Q_p\left(
\left\{
\frac{|x'_j-x^*_j|}
{\max\{|x'_j|,|x^*_j|\}+\epsilon}
\right\}_{j\in\mathcal C}
\right).
\]

\STATE \textbf{Define}
\[
\mathsf{Within}_{\mathcal B^{(s)}}(G'_{b_i},G^*_{b_i})=1
\]
iff, after flattening and restricting to \(\mathcal C\), for every \(p\in\Lambda\),
\[
P_i^{\mathrm{abs}}(p)\le B_s^{\mathrm{abs}}(p)
\;\wedge\;
P_i^{\mathrm{rel}}(p)\le B_s^{\mathrm{rel}}(p).
\]

\RETURN \textsc{accept} iff all checks pass and
\[
\mathsf{Within}_{\mathcal B^{(s)}}(G'_{b_i},G^*_{b_i})=1.
\]

\end{algorithmic}
\vspace{0.25em}
\hrule
\end{minipage}
\end{figure}

\subsection{Mechanism and Settlement}
\label{sec:commitment-settlement}

\mysys{} uses a standard optimistic deposit and slashing settlement layer. Mechanism design is not the main contribution of this paper; mature fraud proof and optimistic verification mechanisms already provide interfaces for commitments, challenge windows, deposits, committee adjudication, and slashing~\cite{kalodner2018arbitrum, ye2024specular, zhao2024takes}. \mysys{} leverages this settlement layer to discourage rational cheating by imposing financial penalties on malicious behaviors. 

Let \(R_{\mathcal U}\) be the reward escrowed by the model owner, \(D_{\mathcal P}\)
the provider deposit, and \(L_{\mathcal P}\le D_{\mathcal P}\) the slash applied
after rejection. Let \(C_{\mathrm{train}}\) be the honest training cost,
\(C_{\mathcal K}(q,n_{\mathcal K})\) the total committee cost for
\(q=\lceil \phi K_s\rceil\) opened intervals and committee size
\(n_{\mathcal K}\), and \(F_{\mathcal K}\) the committee fee. The settlement
layer chooses
\begin{equation}
F_{\mathcal K}\ge C_{\mathcal K}(q,n_{\mathcal K}),
\qquad
L_{\mathcal P}\ge F_{\mathcal K}.
\label{eq:committee-fee-slash}
\end{equation}
If all sampled intervals pass, the provider receives \(R_{\mathcal U}\) and its deposit is returned. If any sampled interval fails, the reward is withheld, the provider is slashed by \(L_{\mathcal P}\), the committee is compensated.

The only honest-provider risk considered here is false rejection. Let
\(\eta_{\mathrm{fr}}\) denote the task level FRR; we use
\(\eta_{\mathrm{fr}}=0.005\) as a conservative deployment value. The expected
honest payoff is
\begin{equation}
\mathbb E[U_{\mathrm{hon}}]
=
(1-\eta_{\mathrm{fr}})R_{\mathcal U}
-
C_{\mathrm{train}}
-
\eta_{\mathrm{fr}}L_{\mathcal P}.
\label{eq:honest-provider-payoff}
\end{equation}

For a deviating provider, let \(C_{\mathrm{dev}}\) be its actual execution cost, \(V_{\mathrm{dev}}\) any additional private value obtained if the deviation finalizes, and \(p_{\mathrm{det}}\) the probability that the randomized audit detects the deviation. Its expected payoff is
\begin{equation}
\mathbb E[U_{\mathrm{dev}}]
=
(1-p_{\mathrm{det}})
(R_{\mathcal U}+V_{\mathrm{dev}})
-
C_{\mathrm{dev}}
-
p_{\mathrm{det}}L_{\mathcal P}.
\label{eq:deviating-provider-payoff}
\end{equation}
Honest execution is preferred when
\begin{equation}
\mathbb E[U_{\mathrm{hon}}]
>
\mathbb E[U_{\mathrm{dev}}],
\label{eq:provider-ic-basic}
\end{equation}
or equivalently,
\begin{equation}
(p_{\mathrm{det}}-\eta_{\mathrm{fr}})
(R_{\mathcal U}+L_{\mathcal P})
>
(C_{\mathrm{train}}-C_{\mathrm{dev}})
+
(1-p_{\mathrm{det}})V_{\mathrm{dev}}.
\label{eq:provider-ic}
\end{equation}
Thus, as long as \(p_{\mathrm{det}}>\eta_{\mathrm{fr}}\), the reward and deposit can be chosen so that deviation has lower expected payoff than honest training. For example, a sufficient deposit satisfies
\begin{equation}
L_{\mathcal P}
>
\frac{
(C_{\mathrm{train}}-C_{\mathrm{dev}})
+
(1-p_{\mathrm{det}})V_{\mathrm{dev}}
}{
p_{\mathrm{det}}-\eta_{\mathrm{fr}}
}
-
R_{\mathcal U}.
\label{eq:sufficient-deposit}
\end{equation}

This economic layer discourages rational cheating by linking detected  attack to financial penalties. Therefore, \(s\), \(\phi\), \(\mathcal B^{(s)}\), and \(D_{\mathcal P}\) should be selected jointly for the target deployment.
\section{Evaluation}
\label{sec:evaluation}

\subsection{Overview}
\label{sec:evaluation-overview}

We evaluate \mysys{} along two dimensions: the empirical gradient boundary and the stride-based audit mechanism. We test whether the boundary accepts honest replay, detects malicious deviations, and remains effective under targeted manipulation, while also measuring the cost and security tradeoff introduced by stride. Unless stated otherwise, stability and FRR experiments use the raw boundary, whereas attack and deployment experiments use the inflated boundary \(\mathcal B^{(s)}\) with \(\alpha_{\mathcal B}=3\).

\subsection{Boundary Calibration and Honest Acceptance}
\label{sec:raw-boundary-validation}
We first measure honest acceptance under heterogeneous replay using the raw boundary. 

We calibrated the raw boundary \(\widetilde{\mathcal B}\) using the heterogeneous GPU set listed in Table~\ref{tab:calibration-gpus}. All experiments used the same AMD EPYC 7352 host CPU. The evaluated workloads are Qwen3-4B on Alpaca, ResNet-152 on CIFAR-10, BERT-large on AG-News, and Stable-Diffusion-v1.5 on Flickr30k.

\begin{table}[t]
\centering
\caption{GPU set used for boundary calibration.}
\label{tab:calibration-gpus}
\papertablefont
\setlength{\tabcolsep}{3pt}
\renewcommand{\arraystretch}{1.05}
\begin{tabularx}{0.8\linewidth}{@{}lLr@{}}
\toprule
\textbf{Vendor} & \textbf{GPU} & \textbf{Memory} \\
\midrule
NVIDIA & L4 & 24 GB \\
NVIDIA & RTX A4500 & 20 GB \\
NVIDIA & RTX 4000 Ada Generation & 20 GB \\
NVIDIA & RTX PRO 4000 Blackwell & 24 GB \\
NVIDIA & RTX PRO 4500 Blackwell & 32 GB \\
NVIDIA & RTX PRO 6000 Blackwell & 96 GB \\
NVIDIA & A100 & 80 GB \\
NVIDIA & GeForce RTX 4090 & 24 GB \\
NVIDIA & GeForce RTX 5090 & 32 GB \\
AMD & Instinct MI300X & 192 GB \\
\bottomrule
\end{tabularx}
\end{table}

\noindent\textbf{Coordinate sampling.}
The following boundary experiments use uniform random coordinate sampling to calibrate and evaluate the percentile boundary. We therefore first measure the effect of sampling on Qwen3-4B, where the checked MLP projection contains about \(24.9\)M gradient coordinates. For each sample size, we draw coordinates uniformly and compare the boundary estimated from sampled coordinates with that computed from all coordinates. 

\begin{figure}[ht]
\centering
\includegraphics[width=0.9\linewidth]{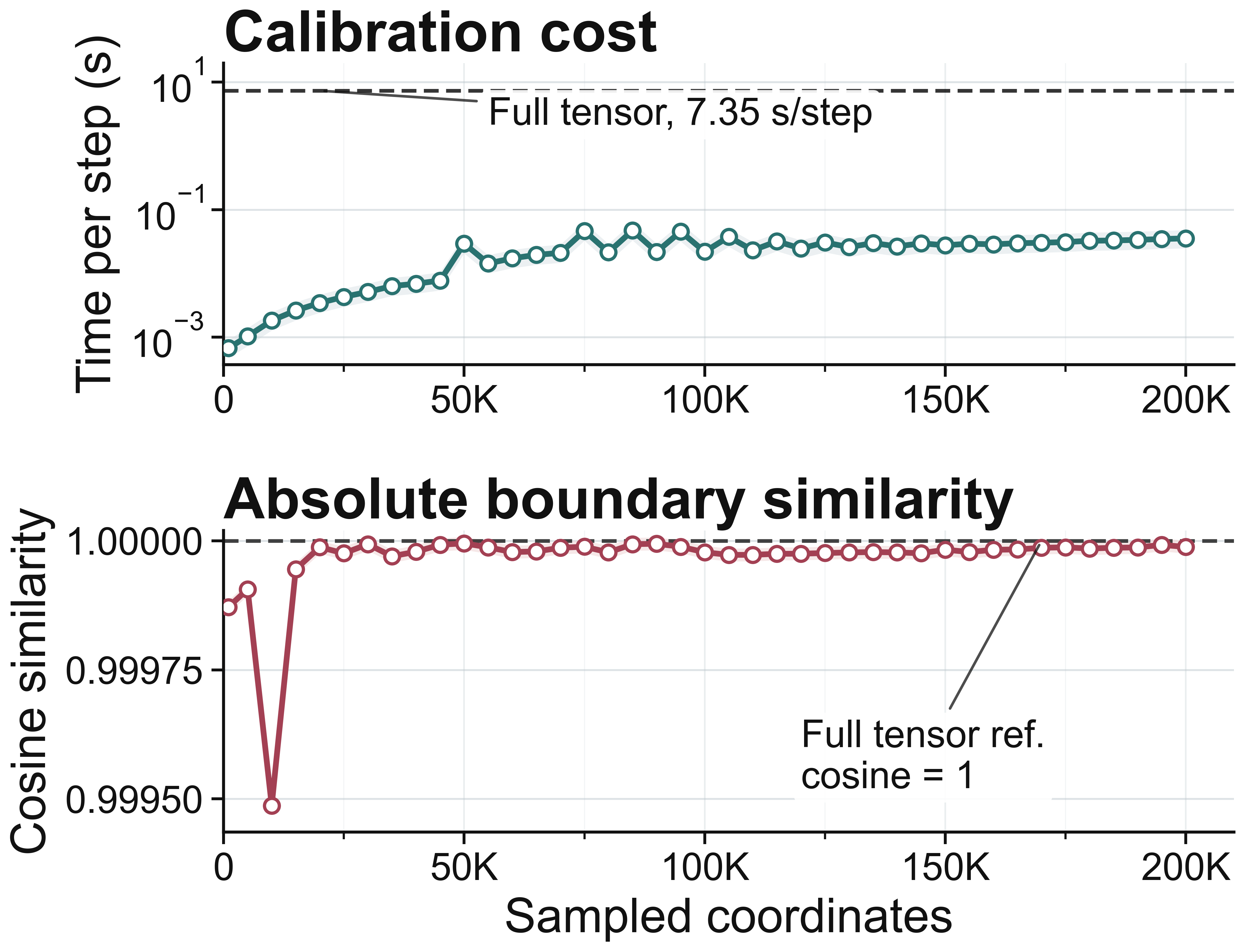}
\caption{Coordinate sampling quality and cost. Panel A: cosine similarity between boundaries from sampled and all coordinates. Panel B: mean percentile computation time per checked step. Dashed lines show results using all coordinates.}
\label{fig:coordinate-sampling-boundary-quality}
\end{figure}

Figure~\ref{fig:coordinate-sampling-boundary-quality} reports both accuracy and efficiency across \(41\) sample sizes. Under the same hardware setting, computing the percentile profile over all coordinates takes \(7.348\) s per checked step. Sampling \(200{,}000\) coordinates reduces this time to \(0.035494\) s, giving a \(207.0\times\) speedup. Compared with the boundary calibrated from full coordinates, the sampling method preserves similarity ranging from \(0.999487\) to \(0.999995\). The finite-population analysis in Appendix~\ref{app:sampling-proof} further bounds the sampled quantile-rank error. Thus, coordinate sampling provides an efficient approximation for the experiments below.

\noindent\textbf{Stability.}
We first evaluate raw boundary stability in the finest-grained setting \(s=1\). For each workload, we calibrate the raw absolute boundary with different numbers of steps and compare its relative drift at percentile coordinates \(\mathcal P=\{30,50,90\}\). Table~\ref{tab:raw-boundary-supnorm} shows that adding more calibration steps does not cause substantial boundary shift. Across all reported workloads and percentile coordinates, the largest relative drift is \(7.3\%\). We further examine stability at \(s=400\) and \(s=200\). In these longer interval settings, early training has higher loss and therefore larger benign deviations, making the calibrated boundary more conservative and stable once early intervals are included. Full diagnostics appear in Appendix~\ref{app:boundary-stability-full}.

\begin{table}[htbp]
\centering
\caption{Raw-boundary stability at selected absolute percentile coordinates.
Values are relative changes, where \(0.01\) corresponds to \(1\%\).}
\label{tab:raw-boundary-supnorm}
\papertablefont
\setlength{\tabcolsep}{4pt}
\renewcommand{\arraystretch}{1.05}
\begin{tabular}{llccc}
\toprule
Model & Steps & $p=30$ & $p=50$ & $p=90$ \\
\midrule
\multirow{4}{*}{BERT-large}
& 100 & 0.013 & 0.042 & 0.000 \\
& 200 & 0.002 & 0.002 & 0.004 \\
& 300 & 0.001 & 0.001 & 0.003 \\
& 400 & 0.001 & 0.001 & 0.003 \\
\midrule
\multirow{4}{*}{Qwen3-4B}
& 100 & 0.000 & 0.001 & 0.001 \\
& 200 & 0.001 & 0.000 & 0.000 \\
& 300 & 0.001 & 0.000 & 0.000 \\
& 400 & 0.001 & 0.000 & 0.000 \\
\midrule
\multirow{4}{*}{ResNet-152}
& 100 & 0.000 & 0.001 & 0.000 \\
& 200 & 0.000 & 0.001 & 0.000 \\
& 300 & 0.000 & 0.001 & 0.000 \\
& 400 & 0.000 & 0.000 & 0.000 \\
\midrule
\multirow{4}{*}{Stable Diffusion}
& 100 & 0.045 & 0.050 & 0.073 \\
& 200 & 0.004 & 0.002 & 0.001 \\
& 300 & 0.001 & 0.001 & 0.001 \\
& 400 & 0.001 & 0.001 & 0.000 \\
\bottomrule
\end{tabular}
\end{table}

\noindent\textbf{Honest false rejection.}
For each workload, we calibrate \(\widetilde{\mathcal B}\) from the first 100 steps and evaluate the following 3000 honest steps. A false rejection occurs when an honest profile exceeds the raw absolute or relative boundary.

\begin{table*}[htbp]
\centering
\caption{FRR of the raw empirical boundary across honest
heterogeneous executions. The boundary is calibrated from the first 100 steps
and evaluated on the following 3000 honest steps.}
\label{tab:raw-boundary-fp-rates}
\setlength{\tabcolsep}{4pt}
\papertablefont
\renewcommand{\arraystretch}{1.05}
\begin{tabularx}{\textwidth}{@{}llLc cc@{}}
\toprule
 \textbf{Model} & \textbf{Dataset} & \textbf{Module} & \textbf{Max Token Length}
& \makecell{\textbf{Absolute FRR}\\\textbf{Rate (\%)}}
& \makecell{\textbf{Relative FRR}\\\textbf{Rate (\%)}} \\
\midrule
Qwen3-4B & Alpaca & layers.35.mlp.down\_proj & 128  & 0.000 & 0.400 \\
Qwen3-4B & Alpaca & layers.35.mlp.down\_proj & 256  & 0.000 & 0.033 \\
Qwen3-4B & Alpaca & layers.35.mlp.down\_proj & 512  & 0.000 & 0.200 \\
Qwen3-4B & Alpaca & layers.35.self\_attn\_q\_proj & 256  & 0.133 & 0.233 \\
ResNet-152 & CIFAR-10 & layer4.2: conv1, conv2, conv3 & N/A  & 0.200 & 0.100 \\
BERT-large & AG-News & bert.encoder.layer.23 & 256  & 0.000 & 0.167 \\
Stable-Diffusion-v1.5 & Flickr30k & mid\_block.attentions.0.transformer\_blocks.0.attn2.to\_k & N/A  & 0.033 & 0.433 \\
\bottomrule
\end{tabularx}
\end{table*}

\begin{figure}[htbp]
\centering
\includegraphics[width=0.95\linewidth]{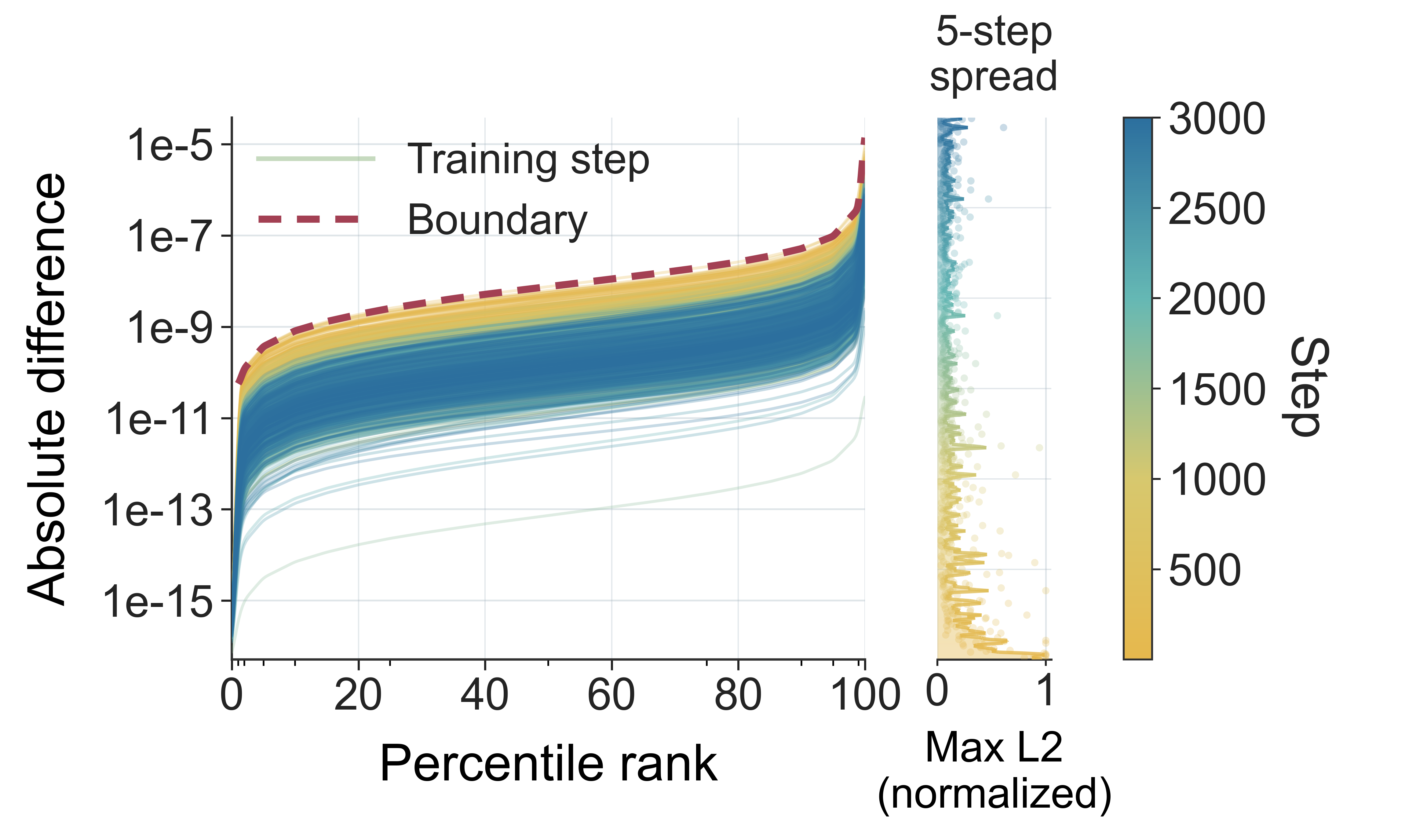}
\caption{Qwen3-4B \(25\)M trainable parameters. Left: absolute percentile deviations vs. the raw boundary. Right: diagnostic normalized maximum L2 distance within each 5-step interval.}
\label{fig:qwen_boundary}
\end{figure}

\begin{figure}[ht]
\centering
\includegraphics[width=0.95\linewidth]{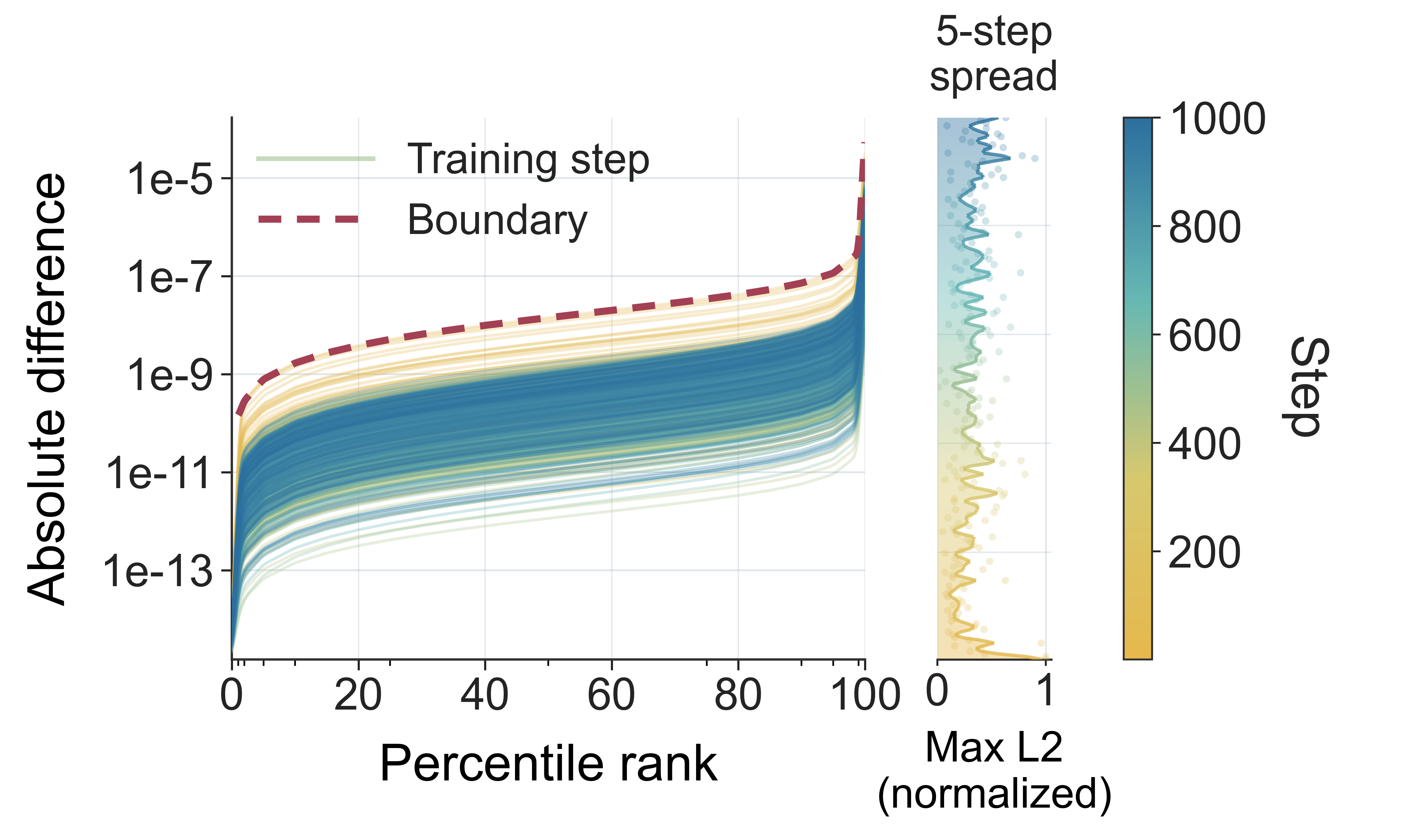}
\caption{Qwen3-4B \(1\)B trainable parameters.}
\label{fig:qwen-1b-boundary}
\end{figure}

Table~\ref{tab:raw-boundary-fp-rates} reports that the maximum FRR is only \(0.433\%\). Figure~\ref{fig:qwen_boundary} visualizes how the gradient difference percentile profiles at individual steps relate to the calibrated boundary. The early steps exhibit larger errors, so using them for calibration yields a conservative boundary. The right-hand ridge plot shows the same pattern: profiles from early training steps have larger normalized L2 distances, while later profiles become more concentrated. When the number of trainable parameters is increased to approximately \(1\)B, the honest profiles still remain below the raw boundary, as shown in Figure~\ref{fig:qwen-1b-boundary}. These results support the use of a small number of early calibration steps to obtain a stable boundary with low FRR.

\noindent\textbf{Robustness of the deployed boundary.}
We evaluate whether a boundary calibrated under one configuration can be reused directly under a different configuration without recalibration. We vary the LoRA layer range, fine-tuning dataset, maximum input length, and verifier GPU. For each source configuration, the boundary is calibrated on 100 honest intervals and then applied to 1,000 disjoint honest intervals from a target configuration using $\alpha_{\mathcal B}=3$. The FRR is at most $0.1\%$ when transferring across LoRA layer ranges or maximum input lengths. For hardware transfer, we calibrate three boundaries pairwise across three GPUs. Each boundary is then used to verify replay traces produced by the other GPU. This yields \(3\times3=9\) replay combinations, with $0\%$ FRR. Dataset transfer shows the largest shift: the NuminaMath-CoT boundary rejects $1.1\%$ of Alpaca intervals, whereas the Alpaca boundary produces no observed false rejections on NuminaMath-CoT. Overall, the calibrated boundary can be reused across the evaluated configurations with low honest rejection rates. Full transfer matrices are reported in Appendix~\ref{app:boundary-robustness}.




\subsection{General Attack Detection}
\label{sec:general-attacks}

This experiment evaluates whether the boundary \(\mathcal B^{(s)}\) rejects gradient-visible deviations while preserving a low FRR. All attack experiments use the inflated boundary \(\mathcal B^{(s)}\) with \(\alpha_{\mathcal B}=3\).
For each workload and each stride \(s\in\{1,100,500,1000,2000,4000\}\), we construct a training trace containing 1,000 stride-aligned intervals and randomly select \(|\mathcal S_{\mathrm{atk}}|=200\) intervals for attack. We record the indices of these attacked intervals. For evaluation, we apply \(\mathsf{VerifyInterval}\) to all 1,000 intervals using the corresponding stride-specific boundary \(\mathcal B^{(s)}\), and record which intervals are rejected. We then compare the attacked and rejected interval sets to measure attack success and false rejection. This evaluates whether \(\mathcal B^{(s)}\) separates attacked intervals from honest intervals.

We evaluate four provider attacks. In micro-batch dropping, the declared batch has \(10\) micro-batches, but the provider uses only \(9\). In stale-update replay, the provider uses a stale gradient computed at an earlier step to form the claimed terminal endpoint. In low-precision execution, it uses an undeclared lower-precision FP16 path and stores the result as declared FP32. In data-path perturbation, it changes the declared input: one non-special neighboring token for Qwen3-4B and BERT, \(10\%\) of normalized spatial locations for ResNet, and \(10\%\) of RGB pixels for Stable-Diffusion.

For \(\mathcal B^{(1)}\), we evaluate the $\mathsf{LogMargin}$ as
\begin{equation}
\mathsf{LogMargin}
=
\frac{1}{|\mathcal{S}_{\mathrm{atk}}|\,|\Lambda|}
\sum_{i\in\mathcal{S}_{\mathrm{atk}}}
\sum_{p\in\Lambda}
\log_{10}
\left(
\frac{P_i^{\mathrm{atk,abs}}(p)}
{B_s^{\mathrm{abs}}(p)}
\right).
\label{eq:attack-log-margin}
\end{equation}
where \(P_i^{\mathrm{atk,abs}}(p)\) is the \(p\)-th percentile of the absolute gradient differences for attacked interval \(i\), and \(B_s^{\mathrm{abs}}(p)\) is the corresponding deployed absolute boundary at stride \(s\). A value of \(3\) means that the geometric mean of \(P_i^{\mathrm{atk,abs}}(p) / B_s^{\mathrm{abs}}(p)\) across attacked intervals and percentile coordinates is \(10^3\).

\begin{table}[htbp]
\footnotesize
\centering
\caption{General attack results. LogMargin is defined in
\cref{eq:attack-log-margin}. }
\label{tab:attack_success}
\papertablefont
\setlength{\tabcolsep}{3.5pt}
\renewcommand{\arraystretch}{1.05}
\begin{tabular}{llccc} 
\toprule
\textbf{Model} & \textbf{Attack Type} & \textbf{ASR} & \textbf{FRR} & \textbf{LogMargin} \\
\midrule
\multirow{4}{*}{Qwen3-4B} & Micro-batch dropping & 0\% & 0\% & 4.595 \\
                           & Stale-update replay  & 0\% & 0\% & 5.472 \\
                           & Low-precision        & 0\% & 0\% & 3.98  \\
                           & Data-path perturbation & 0\% & 0\% & 2.86  \\
\midrule
\multirow{4}{*}{ResNet}   & Micro-batch dropping & 0\% & 0\% & 3.106 \\
                           & Stale-update replay  & 0\% & 0\% & 3.921 \\
                           & Low-precision        & 0\% & 0\% & 4.10  \\
                           & Data-path perturbation & 0\% & 0\% & 4.09  \\
\midrule
\multirow{4}{*}{BERT}     & Micro-batch dropping & 0\% & 0\% & 4.334 \\
                           & Stale-update replay  & 0\% & 0\% & 4.809 \\
                           & Low-precision        & 0\% & 0\% & 3.34  \\
                           & Data-path perturbation & 0\% & 0\% & 3.01  \\
\midrule
\multirow{4}{*}{Stable-Diffusion} & Micro-batch dropping & 0\% & 0\% & 5.251 \\
                           & Stale-update replay  & 0\% & 0\% & 6.378 \\
                           & Low-precision        & 0\% & 0\% & 4.35  \\
                           & Data-path perturbation & 0\% & 0\% & 4.09  \\
\bottomrule
\end{tabular}
\end{table}

Table~\ref{tab:attack_success} shows \(0\%\) ASR for every evaluated attack and workload. The \(\mathsf{LogMargin}\) values range from \(2.86\) to \(6.378\), so these deviations are rejected with margins several orders of magnitude above the calibrated tolerance.

\subsection{Adaptive Target-Directed Gradient Attack}
\label{sec:target-perturbation-attack}
We evaluate an adaptive attack based on PGD to further test \mysys{} and to compare it with baseline methods. The attack directly modifies the update gradient at selected training steps. For each attacked step \(t\), the provider first selects a target token or target label. It then computes two gradients at the same start weight \(W_{t}^{(m)}\), without applying either update. The first is the normal training gradient on the declared step-\(t\) batch \(I_t\):
\begin{equation}
G_t^{\mathrm{normal}}
=
\nabla_{W^{(m)}}\mathcal L_{\mathcal T}
\left(
W_{t}^{(m)}, I_t;\Pi,\Omega
\right).
\label{eq:target-normal-gradient}
\end{equation}
The second is a target-directed gradient computed on a target batch \(I^{\mathrm{target}}\), constructed to contain the chosen target token or target label:
\begin{equation}
G_t^{\mathrm{target}}
=
\nabla_{W^{(m)}}\mathcal L_{\mathcal T}
\left(
W_{t}^{(m)}, I^{\mathrm{target}};\Pi,\Omega
\right).
\label{eq:target-direction-gradient}
\end{equation}
The provider then applies the adversarial gradient update with
\begin{equation}
G_t^{\mathrm{adv}}(\lambda)
=
G_t^{\mathrm{normal}}
+
\lambda G_t^{\mathrm{target}},
\label{eq:target-adv-gradient}
\end{equation}
where \(\lambda\in [0,1]\) controls the attack strength. When \(\lambda=0\), the update is normal; as \(\lambda\) increases, the update becomes more target-directed. We consider each verification method as a \textbf{constraint} on \(\lambda\). The unconstrained setting allows any \(\lambda\), while the constrained branch restricts \(\lambda\). The attack will try to maximize \(\lambda\) subject to the constraint, to achieve the strongest possible target manipulation that still passes verification. For \mysys{}, the maximum \(\lambda\) should satisfy:
\begin{equation}
\lambda_t^\star
=
\max_{\lambda\in [0,1]}
\left\{
\lambda:
\mathsf{Within}_{\mathcal B^{(s)}}
\left(
G'_{b_i},
G^*_{b_i}(\lambda)
\right)=1
\right\}.
\label{eq:target-lambda-constraint}
\end{equation}
In experiments, \(\lambda_t^\star\) is estimated by binary search over a finite search range under deployed boundary \(\mathcal B^{(s)}\). 

For each of the three evaluated models, we construct a separate set of \(100\) target samples. We first train the initial model honestly for \(100\) steps. Using this honestly trained model, we then run inference on \(10{,}000\) held-out samples that were not used in those training steps, retaining only samples that are correctly predicted by the model. For each retained sample, we compute the logit gap, \(\delta_{\mathrm{gap}}\), between the top-1 and top-2 token or label. We rank these samples by \(\delta_{\mathrm{gap}}\) and select the \(100\) lowest samples. For each selected sample, the target is chosen as the top-2 token or label. We form \(100\) target batches \(I^{\mathrm{target}}\) by replacing the attacked token or label in each selected sample with its top-2 alternative. 
Starting again from the initial model, the provider trains for \(100\) steps with the adversarial gradient in \cref{eq:target-adv-gradient}, then evaluates the perturbed model on the selected \(100\) samples to compute the logit increase \(\delta_{\mathrm{inc}}\). We also denote the mean value of \(\delta_{\mathrm{inc}}\) across the \(100\) samples as \(\bar{\delta}_{\mathrm{inc}}\). An attack succeeds when an originally correct prediction is changed to the selected target, equivalently when \(\delta_{\mathrm{inc}}\) crosses the original gap \(\delta_{\mathrm{gap}}\). 

\begin{figure}[!htbp]
\centering
\includegraphics[width=0.4\textwidth]{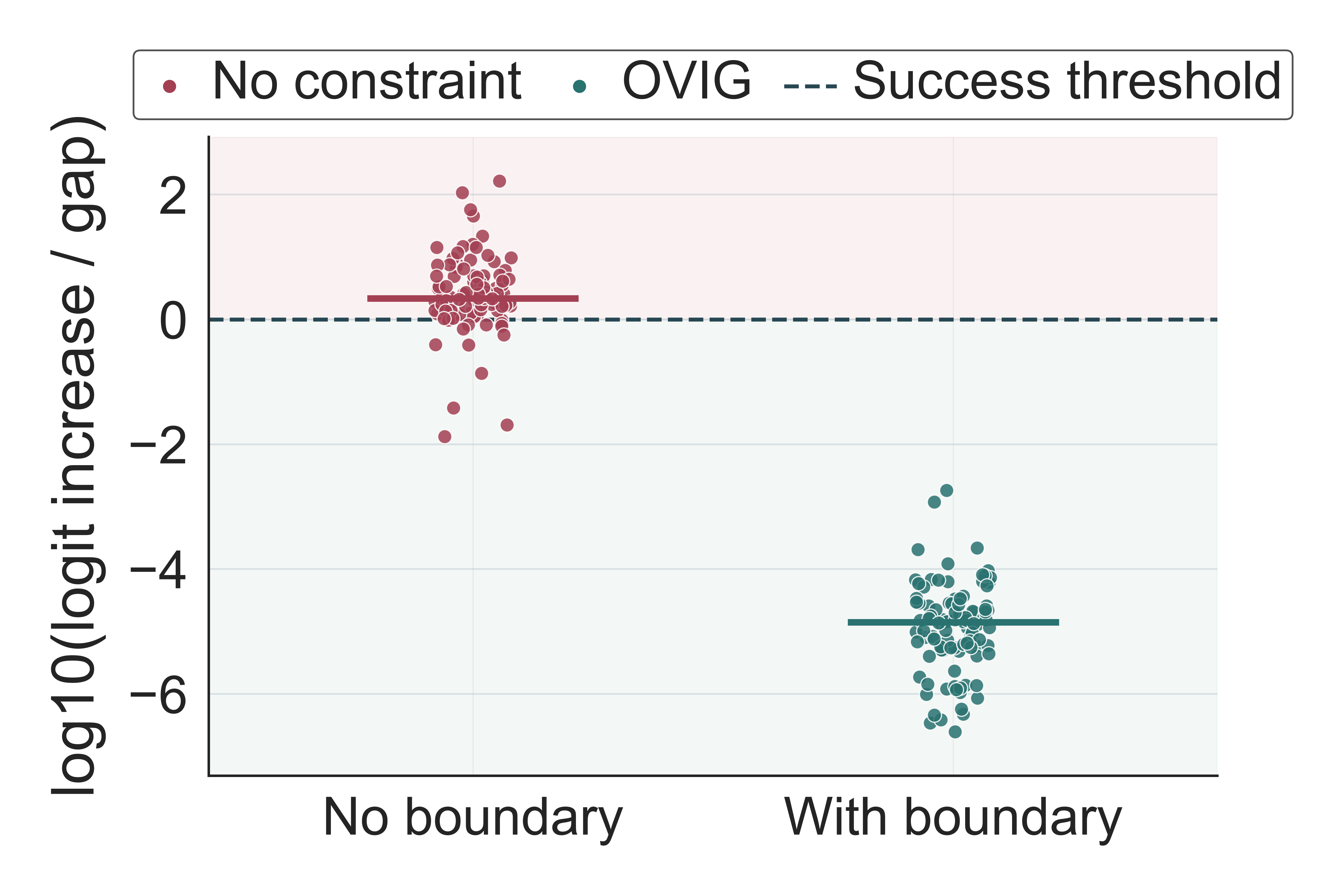}
\caption{Qwen3-4B PGD-based target manipulation. Short horizontal bars show medians. Points above the dashed line indicate successful attacks.}
\label{fig:qwen-target-perturbation}
\end{figure}

Figure~\ref{fig:qwen-target-perturbation} visualizes Qwen3-4B attack outcomes using the gap-normalized log-ratio \(r_{\mathrm{gap}}\) from \cref{eq:gap-normalized-log-ratio}. Without constraints, the attack reaches \(86.0\%\) ASR, with mean target-logit increase \(0.618\) and median \(r_{\mathrm{gap}}=0.339\). Under \mysys{} constraint, ASR drops to \(0.0\%\), \(\bar{\delta}_{\mathrm{inc}}\) falls to \(3.36\times10^{-6}\), and the median \(r_{\mathrm{gap}}\) shifts to \(-4.851\).

\begin{equation}
r_{\mathrm{gap}}
=
\log_{10}
\frac{
\max\{\delta_{\mathrm{inc}},0\}+\epsilon
}{
\delta_{\mathrm{gap}}+\epsilon
}.
\label{eq:gap-normalized-log-ratio}
\end{equation}

Table~\ref{tab:target-perturbation-summary} shows the same pattern for the classification workloads. For both ResNet-152 and BERT-large, the unconstrained attack reaches \(100.0\%\) ASR, whereas the \mysys{} constraint reduces ASR to \(0.0\%\) and lowers the mean target-logit increase by more than four orders of magnitude.

\begin{table}[htbp]
\centering
\caption{PGD-based target manipulation summary. ASR is the fraction of attacks
that cross gap between the original and target logits.}
\label{tab:target-perturbation-summary}
\papertablefont
\setlength{\tabcolsep}{2.4pt}
\renewcommand{\arraystretch}{1.05}
\begin{tabular*}{\linewidth}{@{\extracolsep{\fill}}lcccc@{}}
\toprule
\textbf{Model} & \makecell{\textbf{Unconstrained}\\\textbf{ASR}}
& \makecell{\textbf{\mysys}\\\textbf{ASR}}
& \makecell{\textbf{Mean $\boldsymbol{\bar{\delta}_{\mathrm{inc}}}$}\\\textbf{(\mysys)}}
& \makecell{\textbf{Mean $\boldsymbol{\bar{\delta}_{\mathrm{inc}}}$}\\\textbf{(Unconstrained)}} \\
\midrule
Qwen3-4B & 86.0\% & 0.0\% & $3.36{\times}10^{-6}$ & $0.618$ \\
ResNet-152 & 100.0\% & 0.0\% & $4.43{\times}10^{-6}$ & $2.47$ \\
BERT-large & 100.0\% & 0.0\% & $1.35{\times}10^{-5}$ & $0.219$ \\
\bottomrule
\end{tabular*}
\end{table}

These results show that \mysys \ suppresses the target direction by orders of magnitude. 

\subsection{Benchmark Checks Lack Sensitivity}
\label{sec:benchmark-level-evidence}

The benchmark experiment asks whether the same targeted manipulation is visible from final model metrics alone. For ResNet-152 and Qwen3-4B, we compare normal and attacked runs under the same 10,000-step schedule. The attacker chooses 10 target samples and manipulates one randomly selected 100-step subset per target. All \(10\) targets succeed.

\begin{table}[htbp]
\centering
\caption{Benchmark metrics under target-directed attack. Accuracies are class-level for ResNet and token-level for Qwen3-4B.}
\label{tab:benchmark_results}
\papertablefont
\setlength{\tabcolsep}{2.2pt}
\renewcommand{\arraystretch}{1.05}
\begin{tabularx}{\linewidth}{@{}lLrrrc@{}}
\toprule
\textbf{Model} & \textbf{Metric} & \textbf{Normal} & \textbf{Attack} &
\textbf{Difference} & \makecell{\textbf{Target}\\\textbf{Success}} \\
\midrule
ResNet & Top-1 accuracy (\%) & 93.140 & 93.060 & $-0.080$ & \multirow{3}{*}{10/10} \\
ResNet & Top-5 accuracy (\%) & 99.790 & 99.790 & 0.000 & \\
ResNet & Loss & 0.203 & 0.204 & $+0.001$ & \\
\midrule
Qwen3-4B & Top-1 accuracy (\%) & 91.335 & 91.358 & $+0.024$ & \multirow{3}{*}{10/10} \\
Qwen3-4B & Top-5 accuracy (\%) & 96.960 & 96.956 & $-0.004$ & \\
Qwen3-4B & Sequence match (\%) & 4.500 & 4.000 & $-0.500$ & \\
\bottomrule
\end{tabularx}
\end{table}

Table~\ref{tab:benchmark_results} shows that aggregate task metrics remain plausible in this evaluated attack setting. This indicates that benchmark-level evidence is too coarse to reliably expose PGD-based target manipulation.

\subsection{Baseline Verifier Methods}
\label{sec:baseline-comparison}

We compare different baseline methods by treating them as constraints for the attack scale \(\lambda\) in \cref{eq:target-adv-gradient}. This is a comparison of verification methods, not a full deployment reproduction of every baseline system. For each method, we find the largest \(\lambda\) within the constraint by binary search over a bounded interval. 

\begin{table}[htbp]
\centering
\caption{Attack effect under different verifier constraints. $\lambda^\star$ is reported
by order of magnitude. $\bar{\delta}_{\mathrm{inc}}$ is defined in \secref{sec:target-perturbation-attack}.}
\label{tab:scaler_comparison}
\papertablefont
\setlength{\tabcolsep}{3.4pt}
\renewcommand{\arraystretch}{1.05}
\begin{tabular}{@{} l c c c @{}}
\toprule
\textbf{Method} & \makecell{\textbf{Admissible}\\\(\boldsymbol{\lambda^\star}\)\textbf{ order}} & \makecell{$\boldsymbol{\bar{\delta}_{\mathrm{inc}}}$} & \textbf{ASR (\%)} \\
\midrule
PoTD & $10^{0}$ & $2.90{\times}10^{-1}$ & 86 \\
RTTD & $10^{-1}$ & $1.00{\times}10^{-2}$ & 64 \\
PoL-L2($s=1$) & $10^{-4}$ & $1.31{\times}10^{-4}$ & 0 \\
\mysys($s=1$) & $10^{-6}$ & $1.45{\times}10^{-6}$ & 0 \\
\bottomrule
\end{tabular}
\end{table}

\begin{table*}[htbp]
\centering
\caption{Attack success rate (ASR) under replay stride. Values are percentages.}
\label{tab:attack_summary}
\papertablefont
\setlength{\tabcolsep}{2.7pt}
\renewcommand{\arraystretch}{1.05}
\begin{tabular*}{0.9\textwidth}{@{\extracolsep{\fill}}ll*{9}{r}@{}}
\toprule
 &  & \multicolumn{8}{c}{Replay stride} \\
\cmidrule(lr){3-11}
\textbf{Model} & \textbf{Method}
& \textbf{1} & \textbf{5} & \textbf{10} & \textbf{20}
& \textbf{50} & \textbf{100} & \textbf{200} & \textbf{400} & \textbf{2000}\\
\midrule
BERT & \mysys
& \safe{0.0\%} & \safe{0.0\%} & \safe{0.0\%} & \safe{0.0\%}
& \safe{0.0\%} & \safe{0.0\%} & \safe{0.0\%} & \safe{0.0\%} & \safe{0.0\%} \\
BERT & PoL-L2
& \unsafe{2.0\%} & \unsafe{2.0\%} & \unsafe{4.0\%} & \unsafe{4.0\%}
& \unsafe{5.0\%} & \unsafe{5.0\%} & \unsafe{10.0\%} & \unsafe{11.0\%} & \unsafe{11.0\%} \\
Qwen3-4B & \mysys
& \safe{0.0\%} & \safe{0.0\%} & \safe{0.0\%} & \safe{0.0\%}
& \safe{0.0\%} & \safe{0.0\%} & \safe{0.0\%} & \safe{0.0\%} & \safe{0.0\%} \\
Qwen3-4B & PoL-L2
& \unsafe{0.0\%} & \unsafe{0.0\%} & \unsafe{5.0\%} & \unsafe{7.0\%}
& \unsafe{10.0\%} & \unsafe{13.0\%} & \unsafe{19.0\%} & \unsafe{31.0\%} & \unsafe{100.0\%}\\
ResNet & \mysys
& \safe{0.0\%} & \safe{0.0\%} & \safe{0.0\%} & \safe{0.0\%}
& \safe{0.0\%} & \safe{0.0\%} & \safe{0.0\%} & \safe{0.0\%} & \safe{0.0\%} \\
ResNet & PoL-L2
& \unsafe{60.0\%} & \unsafe{97.0\%} & \unsafe{100.0\%} & \unsafe{100.0\%}
& \unsafe{100.0\%} & \unsafe{100.0\%} & \unsafe{100.0\%} & \unsafe{100.0\%} & \unsafe{100.0\%}\\
\bottomrule
\end{tabular*}
\end{table*}

The compared constraints observe different evidence. PoL constrains replay deviation with a \(p\)-norm distance between endpoints; in our experiments, we instantiate this family with the L2 norm and refer to it as PoL-L2. PoL-L2 uses the calibration setting of \cref{sec:protocol-boundary-calibration}, but constrains \(\lambda\) by the maximum endpoint L2 distance~\cite{jia2021proof}. PoTD is instantiated as a \(10\)-step segment-level memorization constraint: a target-manipulation run is admissible only if its segment score remains below the normal transcript threshold \(\gamma_{\mathrm{PoTD}}\)~\cite{choi2023tools}. RTTD is instantiated with \(7\) servers. One server performs training with the PGD-based target manipulation in \cref{eq:target-adv-gradient}, while the other \(6\) servers run honest replicas from the same checkpoint. Their honest executions form an aggregate Zest/KS-style cluster, which constrains \(\lambda\) by testing whether the attacked execution remains behaviorally consistent with the cluster~\cite{jia2025backdoor}. We use \(0.01\) as the rejection significance level. Exact score definitions and calibration procedures appear in Appendix~\ref{app:baseline-details}.

Table~\ref{tab:scaler_comparison} shows the PGD-based target manipulation attack effect for Qwen3-4B under the different constraints. PoL-L2 and \mysys \  yield \(0\%\) ASR at \(s=1\), but \mysys \ admits a smaller perturbation scale. PoTD and RTTD are attackable, with ASR \(86\%\) and \(64\%\).

We use PoL-L2 for the stride sweep because it is more directly comparable to OVIG: both constrain replay consistency using endpoint evidence. 

\begin{figure}[htbp]
\centering
\includegraphics[width=0.95\linewidth]{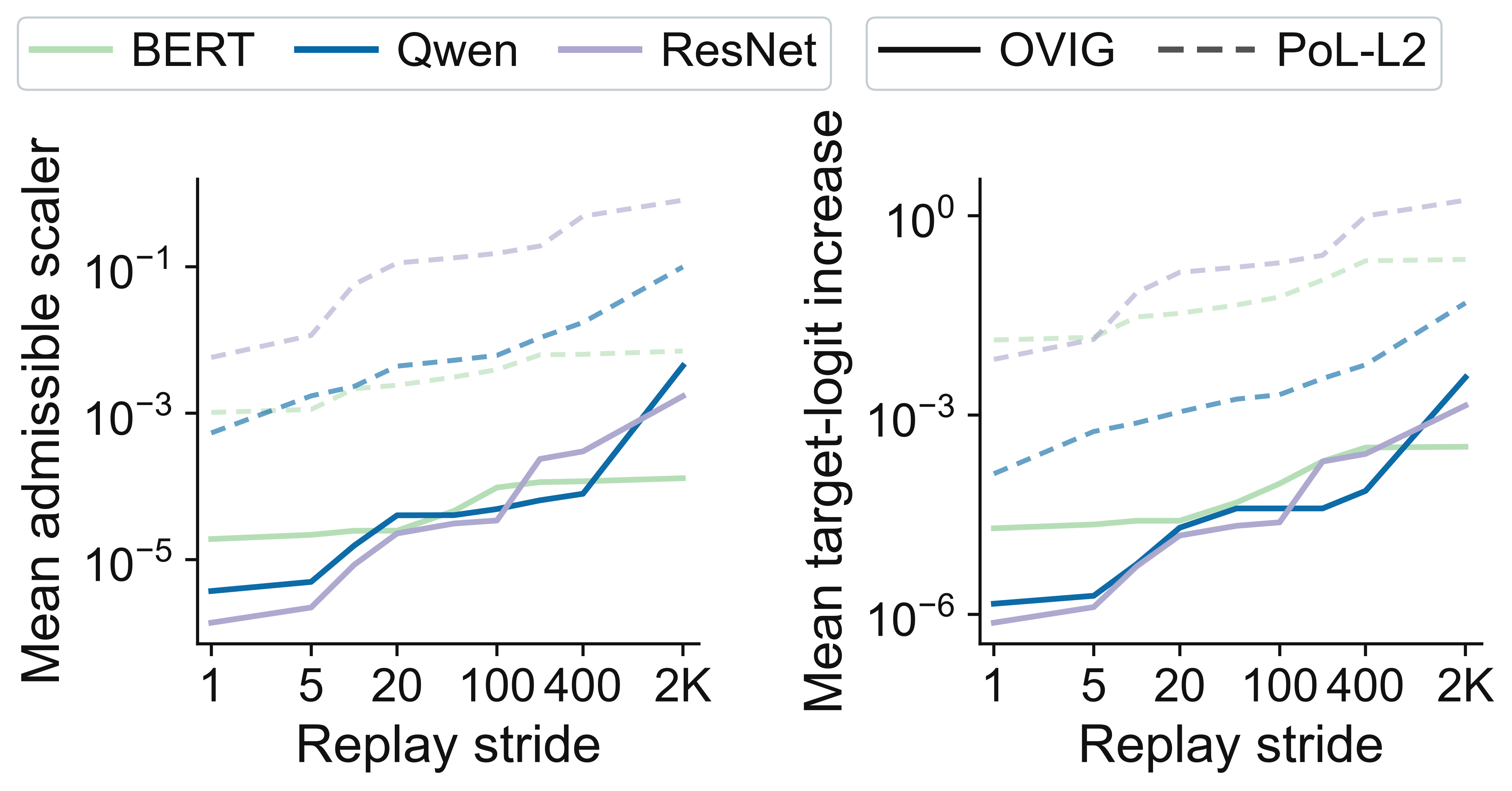}
\caption{Baseline comparison under replay stride. Top: mean admissible attack scale. Bottom: mean target-logit increase induced by the admitted perturbation.}
\label{fig:baseline-panels-bc}
\end{figure}

Table~\ref{tab:attack_summary} and Figure~\ref{fig:baseline-panels-bc} show that \mysys{} remains at \(0.0\%\) ASR across all evaluated strides and models. PoL-L2 becomes increasingly permissive as stride grows: BERT reaches \(11.0\%\) ASR and Qwen3-4B reaches \(100.0\%\) ASR at \(s=2000\), while ResNet reaches \(100.0\%\) ASR for \(s\ge10\). These results motivate the stride-cost analysis in the next section.

\subsection{Stride-Based Deployment and Cost}
\label{sec:consumption-tradeoff}

We evaluate the system cost and required provider deposit under the sampling rules in \secref{sec:protocol-design}, with verification following Algorithm~\ref{alg:stride-audit}. We fine-tune the $24.9$M-parameter MLP projection of Qwen3-4B using SGD and model a training run of \(N=10^6\) steps. We use a five-member committee and evaluate fixed audit fractions. Deposit requirements follow the settlement model in \secref{sec:commitment-settlement}. All compute measurements are taken on an RTX 4090 GPU with an AMD EPYC 7352 CPU, and monetary costs are estimated using the GPU rental rate on RunPod.

\noindent\textbf{System overhead over unverified training.} Table~\ref{tab:system-overhead-average-case} reports compute overhead relative to the same \(N\)-step training run without verification. The accounting includes constructing the final-model commitment \(\mathsf{MR}(\widehat{\mathcal M}_N)\) and the endpoint-state commitment \(R_{\mathrm{state}}\), as defined in \cref{eq:provider-commitments}, as well as committee verification for \(q\) opened intervals. It excludes one-time boundary calibration and task publication. 

\begin{table}[t]
\centering
\caption{System overhead relative to unverified training.}
\label{tab:system-overhead-average-case}
\papertablefont
\begin{tabular}{rrrr}
\toprule
$\phi$ & Stride $s$ & Committee verification (h) & Total / unverified \\
\midrule
0.01 & 1 & 67.35 & 8.209$\times$ \\
0.01 & 100 & 1.55 & 1.100$\times$ \\
0.01 & 500 & 1.01 & 1.042$\times$ \\
0.01 & 1000 & 0.95 & 1.035$\times$ \\
0.01 & 2000 & 0.92 & 1.031$\times$ \\
\addlinespace
0.05 & 1 & 336.75 & 16.469$\times$ \\
0.05 & 100 & 7.73 & 1.289$\times$ \\
0.05 & 500 & 5.07 & 1.167$\times$ \\
0.05 & 1000 & 4.74 & 1.151$\times$ \\
0.05 & 2000 & 4.58 & 1.143$\times$ \\
\addlinespace
0.10 & 1 & 673.50 & 26.795$\times$ \\
0.10 & 100 & 15.47 & 1.526$\times$ \\
0.10 & 500 & 10.15 & 1.322$\times$ \\
0.10 & 1000 & 9.48 & 1.297$\times$ \\
0.10 & 2000 & 9.15 & 1.284$\times$ \\
\addlinespace
0.15 & 1 & 1,010 & 37.121$\times$ \\
0.15 & 100 & 23.20 & 1.764$\times$ \\
0.15 & 500 & 15.22 & 1.478$\times$ \\
0.15 & 1000 & 14.23 & 1.442$\times$ \\
0.15 & 2000 & 13.73 & 1.424$\times$ \\
\bottomrule
\end{tabular}
\end{table}

Table~\ref{tab:system-overhead-average-case} shows that system overhead decreases as \(s\) increases and \(\phi\) decreases. Committee verification time includes replay computation, loading the endpoint states, verifying Merkle authentication paths, reconstructing the replay inputs, and computing the endpoint gradients and percentile profiles.
For a fixed \(\phi\), the expected total replayed training steps are comparable across strides, but smaller strides require the committee to open more intervals and repeatedly load checkpoints and authenticate Merkle paths. Larger strides reduce such overhead.

\noindent\textbf{Audit probability and deposit.}
Although Table~\ref{tab:system-overhead-average-case} shows that smaller \(\phi\) reduces system overhead, the audit fraction cannot be driven arbitrarily low. By the sufficient deposit condition in \cref{eq:sufficient-deposit}, lowering the detection probability increases provider's deposit. Table~\ref{tab:phi-stride-deposit-tradeoff-compact} therefore reports representative \((\phi,s)\) choices for the task setting in this section. The appropriate \(\phi\) should be chosen jointly with the target detection probability and acceptable deposit.

\begin{table}[t]
\centering
\caption{Tradeoff between \(\phi, s\), and deposit. Costs are in USD.}
\label{tab:phi-stride-deposit-tradeoff-compact}
\papertablefont
\begin{tabular}{rrrrr}
\toprule
\(\phi\) & Stride & \(q\) & Committee cost & Required deposit \\
\midrule
0.01 & 1 & 10,000 & 47.08 & 2,481.52 \\
0.01 & 500 & 20 & 1.31 & 2,435.75 \\
0.01 & 1000 & 10 & 1.26 & 2,435.70 \\
0.01 & 2000 & 5 & 1.24 & 2,435.68 \\
\addlinespace
0.05 & 1 & 50,000 & 235.40 & 722.29 \\
0.05 & 500 & 100 & 6.54 & 493.43 \\
0.05 & 1000 & 50 & 6.31 & 493.20 \\
0.05 & 2000 & 25 & 6.20 & 493.09 \\
\addlinespace
0.1 & 1 & 100,000 & 470.80 & 714.24 \\
0.1 & 500 & 200 & 13.09 & 256.53 \\
0.1 & 1000 & 100 & 12.63 & 256.07 \\
0.1 & 2000 & 50 & 12.40 & 255.84 \\
\addlinespace
0.15 & 1 & 150,000 & 706.20 & 868.49 \\
0.15 & 500 & 300 & 19.63 & 181.93 \\
0.15 & 1000 & 150 & 18.94 & 181.24 \\
0.15 & 2000 & 75 & 18.60 & 180.90 \\
\bottomrule
\end{tabular}
\end{table}

The required deposit in Table~\ref{tab:phi-stride-deposit-tradeoff-compact} also accounts for the blockchain commitment cost, measured on the Base Sepolia testnet. With Merkle root commitments, this cost does not scale with stride \(s\); the randomized audit path remains nearly constant at about \(801{,}833\) gas. Combining the overhead and deposit results, \(\phi=0.05\) or \(\phi=0.10\) with \(s=2000\) provides a practical operating point for this task setting.

\noindent\textbf{Off-chain storage and evidence transmission.}
Off-chain storage is the provider-side evidence retained after training. It includes the stride-aligned replay evidence $\mathcal E_i$, consisting of the trainable parameters, replay metadata, and, when required, optimizer state at the interval boundary. Evidence transmission is the data sent to each committee member when an interval is opened, including the opened endpoints, per-step replay metadata, data authentication metadata, and Merkle authentication paths. Since \(K_s=\lceil N/s\rceil\), increasing \(s\) reduces the number of retained endpoint records and the expected number of interval openings.

\begin{figure}[htbp]
\centering
\includegraphics[width=0.95\linewidth]{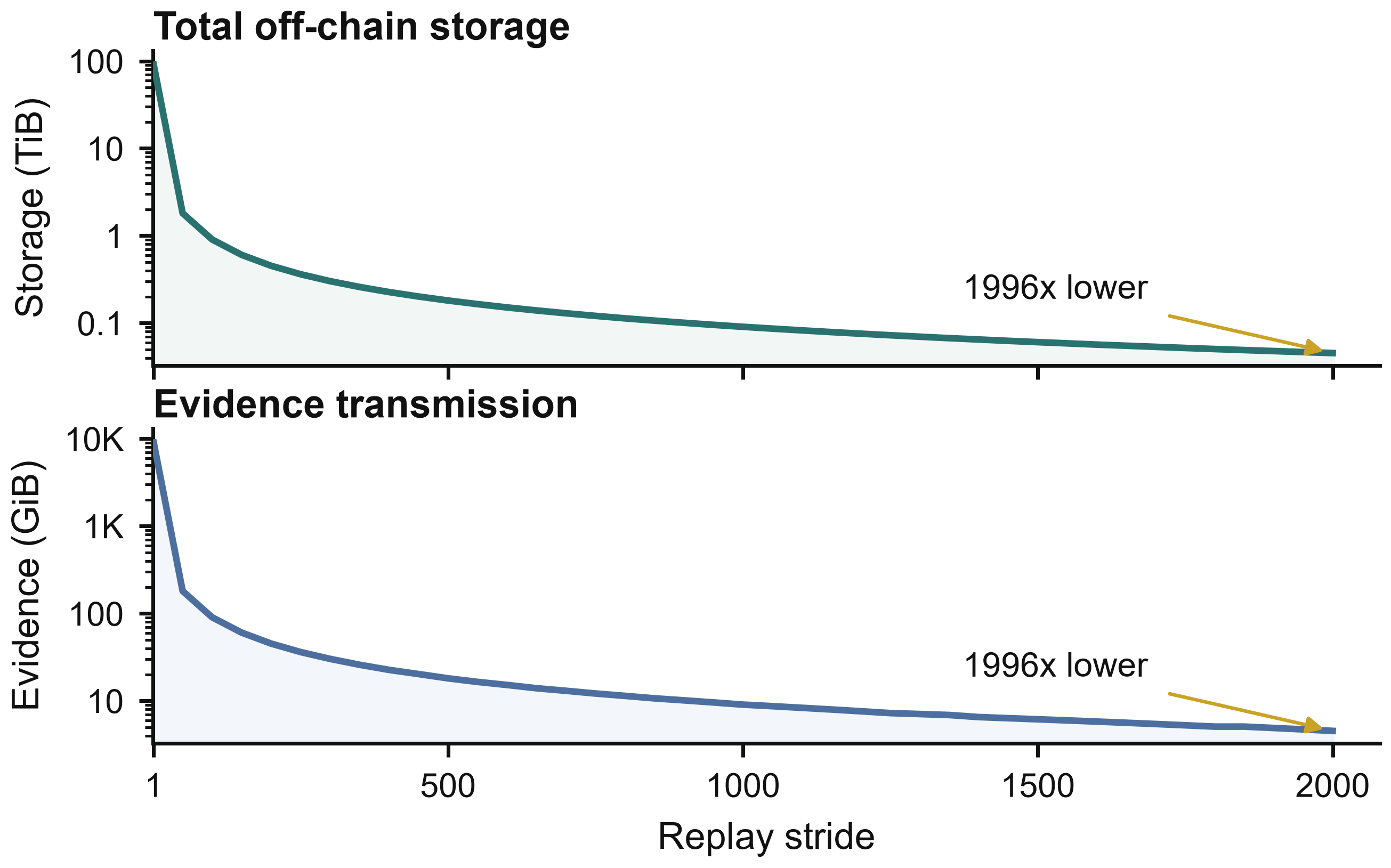}
\caption{Stride effects on off-chain storage and evidence transmission. Top: total off-chain storage in TiB. Bottom: expected evidence sent to each committee member at \(\phi=0.05\), in GiB.}
\label{fig:stride-storage-transmission}
\end{figure}

Figure~\ref{fig:stride-storage-transmission} reports the scaling with stride of off-chain storage and audit-evidence transmission. Increasing \(s\) from \(1\) to \(2000\) reduces storage from \(90.60\) TiB to \(46.48\) GiB, and reduces the expected audit evidence sent to each committee member at \(\phi=0.05\) from \(8.83\) TiB to \(4.53\) GiB. Both reductions are approximately \(1996\times\).

We also compare \mysys~with~\cite{srivastava2024optimisticDeterministic} under the same deployment setting. That approach controls nondeterminism by recording the floating point calculation rounding decisions. Even when only a subset of parameters is trained, deterministic replay still requires recording decisions along the full model execution path, resulting in substantial overhead. In this accounting comparison, its verification overhead is \(4.75\times\) larger than \mysys{}'s, and its off-chain storage and evidence transmission are \(1148.94\times\) larger.

\noindent\textbf{Stride selection.}
In practice, the stride should be chosen as the largest value that satisfies the deployment security requirements. Let $\mathcal S$ denote the candidate stride set, $\eta_{\mathrm{fr}}(s)$ the honest false-rejection rate under $\mathcal B^{(s)}$, and $A_{\mathrm{adv}}(s)$ the strongest adaptive attack effect admitted by the same boundary. The deployment specifies the maximum acceptable values $\eta_{\max}$ and $A_{\max}$. We then select
\begin{equation}
s^\star
=
\max_{s\in\mathcal S}
\left\{
s:
\eta_{\mathrm{fr}}(s)\le \eta_{\max},
\;
A_{\mathrm{adv}}(s)\le A_{\max}
\right\}.
\label{eq:stride-selection}
\end{equation}
Among strides satisfying these security constraints, a larger $s$ is preferred because it reduces the number of retained endpoints and audit openings. In our evaluated setting, an admissible attack scale $\lambda^\star$ on the order of $10^{-4}$ or smaller provides a practical reference for the attack constraint in \secref{sec:baseline-comparison}. The audit fraction $\phi$ can be selected according to the deposit amount acceptable to the provider.

\section{Limitations and Conclusion}
\label{sec:limitations-conclusion}

\noindent\textbf{Limitations.}
\mysys{} relies on an empirically calibrated boundary. Our experiments show that this boundary is stable, achieves low FRR, and transfers well across the evaluated settings. However, we cannot exhaustively cover all possible training configurations, so a boundary calibrated under one setting may not remain valid after substantial changes to the model, training policy, numerical stack, or hardware. In addition, the security of \mysys{} depends on the settlement mechanism, the assumption of rational participants, and a committee with an honest majority, rather than on a universal proof of correct execution. Finally, \mysys{} does not provide data privacy during audits; incorporating privacy preserving verification is left for future work.

\noindent\textbf{Conclusion.}
We presented OVIG, an optimistic protocol for verifying outsourced AI training under heterogeneous execution. OVIG binds the returned model to a stride-aligned endpoint trajectory and audits sampled intervals by replay and comparing gradient differences against an empirically calibrated percentile boundary. Across the evaluated language, vision, and diffusion workloads, OVIG detected all tested shortcut and target-directed deviations in opened intervals, while substantially reducing evidence storage and transmission. These results demonstrate that gradient signals provide a practical integrity check without requiring bitwise deterministic replay or heavyweight cryptographic proofs.

\bibliographystyle{IEEEtran}
\bibliography{references}

\appendices
\section{Rationale for the Gradient-Percentile Metric}
\label{app:metric-rationale}

This appendix formalizes the two observations motivating the verifier signal used by OVIG: endpoint-gradient discrepancies incorporate the local geometry of the declared training objective, while a percentile profile preserves more of the coordinate-error distribution than a single scalar norm.

\noindent\textbf{Gradient discrepancy and local loss geometry.}
Let
\[
g_I(W)=\nabla_W L(W;I)
\]
denote the gradient of the declared loss on the fixed endpoint-check batch $I$.  Consider two endpoints $W$ and $W+\delta W$.

\begin{proposition}
If $L(\cdot;I)$ is twice continuously differentiable along the segment between
$W$ and $W+\delta W$, then
\[
g_I(W+\delta W)-g_I(W)
=
\left(
\int_0^1 H_I(W+t\delta W)\,dt
\right)\delta W,
\]
where $H_I(W)=\nabla_W^2 L(W;I)$.
\end{proposition}

\begin{IEEEproof}
Define $h(t)=g_I(W+t\delta W)$.  By the chain rule,
\[
h'(t)=H_I(W+t\delta W)\delta W.
\]
Applying the fundamental theorem of calculus gives
\[
h(1)-h(0)
=
\int_0^1 H_I(W+t\delta W)\delta W\,dt,
\]
which proves the result.
\end{IEEEproof}

Thus, a weight-space norm captures only the magnitude of endpoint displacement, whereas gradient discrepancy also reflects the local loss geometry. Under a locally quadratic loss, equal-norm perturbations along Hessian eigenvectors \(v_1\) and \(v_2\) induce discrepancies \(r|\lambda_1|\) and \(r|\lambda_2|\), respectively, allowing OVIG to assess deviations through the declared training objective.

\noindent\textbf{Percentile profiles and scalar norms.}
Let
\[
z_j = |g'_j-g^\star_j|,\qquad j=1,\ldots,d,
\]
be the coordinate-wise absolute gradient-error vector, and let $Q_z(u)$ be its empirical quantile function.

\begin{proposition}
For any $q\geq 1$,
\[
\frac{1}{d}\|z\|_q^q
=
\int_0^1 Q_z(u)^q\,du.
\]
\end{proposition}

\begin{IEEEproof}
Let
$z_{(1)}\leq\cdots\leq z_{(d)}$
be the ordered coordinates.  The empirical quantile function is constant at $z_{(j)}$ over the corresponding interval of width $1/d$.  Therefore,
\[
\int_0^1 Q_z(u)^q\,du
=
\frac{1}{d}\sum_{j=1}^d z_{(j)}^q
=
\frac{1}{d}\|z\|_q^q.
\]
\end{IEEEproof}

Hence, an $\ell_q$ norm is a single aggregate functional of the quantile curve and does not characterize how errors are distributed across coordinates. OVIG instead evaluates a fixed percentile grid $\{Q_z(p):p\in\Lambda\}$, preserving information about both typical and tail deviations with only $O(|\Lambda|)$ values rather than the $O(d)$ values required by the full quantile profile.
\section{Sampling Bound for Fixed Coordinate Calibration}
\label{app:sampling-proof}

This appendix gives the finite-population argument behind fixed-coordinate calibration. The coordinate set is not an infinite distribution: it is the finite population of gradient coordinates in the checked module. Sampling coordinates uniformly without replacement therefore admits a direct Hoeffding--Serfling concentration bound. The result is a rank-space guarantee: sampled percentiles are close to the full-coordinate percentiles in percentile level, which is the quantity used by the empirical boundary.

\subsection{Finite-Population CDF Concentration}

Let $z_1,\ldots,z_d$ be the coordinate-level values whose percentile profile is being estimated, such as absolute cross-device spreads for one calibration step. A verifier samples $n<d$ coordinates uniformly without replacement and uses the sampled values to estimate the full-coordinate percentile profile. For any threshold $x$, define the indicator population
\begin{equation}
Y_i^{(x)}=\mathbf{1}_{\{z_i\leq x\}}, \qquad i=1,\ldots,d.
\end{equation}
The full-coordinate and sampled empirical CDFs are
\begin{equation}
F_d(x)=\frac{1}{d}\sum_{i=1}^{d}Y_i^{(x)}, \qquad
F_n(x)=\frac{1}{n}\sum_{i=1}^{n}Y_i^{(x)}.
\end{equation}
Since each $Y_i^{(x)}$ lies in $[0,1]$, the Hoeffding--Serfling inequality for sampling without replacement~\cite{bardenet2015concentration} gives, for every fixed $x$ and every $\varepsilon>0$,
\begin{equation}
\Pr\!\left(\left|F_n(x)-F_d(x)\right|\geq\varepsilon\right)
\leq
2\exp\!\left(
-\frac{2n\varepsilon^2}{1-(n-1)/d}
\right).
\label{eq:sampling-cdf-point}
\end{equation}
The denominator is the finite-population correction. It is slightly smaller than the infinite-population Hoeffding denominator, so ignoring it gives a conservative but nearly identical bound when $d$ is large.

For a finite set of $L$ threshold values needed to evaluate the boundary percentile grid, a union bound yields
\begin{align}
&\Pr\!\left(
\max_{1\leq \ell\leq L}\left|F_n(t_{\ell})-F_d(t_{\ell})\right|
\geq \varepsilon
\right) \nonumber\\
&\qquad \leq
2L\exp\!\left(
-\frac{2n\varepsilon^2}{1-(n-1)/d}
\right).
\label{eq:sampling-grid-bound}
\end{align}
Thus, with probability at least $1-\alpha$, the sampled CDF is within
$\varepsilon$ of the full-coordinate CDF on the evaluated grid whenever
\begin{equation}
\varepsilon
=
\sqrt{
\left(\frac{1}{2n}-\frac{n-1}{2nd}\right)
\ln\!\left(\frac{2L}{\alpha}\right)
}.
\label{eq:sampling-epsilon}
\end{equation}

\subsection{From CDF Error to Quantile Error}

Let the full-coordinate and sampled quantile functions be
\begin{equation}
\begin{aligned}
Q_d(u)&=\inf\{x:F_d(x)\geq u\},\\
Q_n(u)&=\inf\{x:F_n(x)\geq u\}.
\end{aligned}
\end{equation}
If the CDF deviation is at most $\varepsilon$ on the relevant threshold range,
then for every quantile level $u\in[\varepsilon,1-\varepsilon]$,
\begin{equation}
Q_d(u-\varepsilon)\leq Q_n(u)\leq Q_d(u+\varepsilon).
\label{eq:sampling-quantile-sandwich}
\end{equation}
To see the right inequality, set $x_+=Q_d(u+\varepsilon)$. Then $F_d(x_+)\geq u+\varepsilon$, so $F_n(x_+)\geq u$ under the CDF deviation event. By the definition of $Q_n$, $Q_n(u)\leq x_+$. For the left inequality, any $x<Q_d(u-\varepsilon)$ satisfies $F_d(x)<u-\varepsilon$, hence $F_n(x)<u$. Therefore $Q_n(u)$ cannot lie below $Q_d(u-\varepsilon)$.

Equation~\eqref{eq:sampling-quantile-sandwich} is the key guarantee for boundary calibration. Coordinate sampling need not reproduce every coordinate value exactly. It only needs to preserve the percentile profile used by the boundary, and the sampled quantile at level $u$ is trapped between neighboring full-coordinate quantiles at levels $u-\varepsilon$ and $u+\varepsilon$.

\subsection{Numerical Scale for the Qwen3-4B Calibration}

For the Qwen3-4B checked projection used in the calibration-cost experiment, $d\approx 24.9$ million coordinates. Using the rounded value $d=25{,}000{,}000$, a percentile grid of size $L=23$, and confidence $1-\alpha=99\%$, \cref{eq:sampling-epsilon} gives
\begin{equation}
\varepsilon \approx 0.00648
\quad\text{for } n=100{,}000,
\end{equation}
and
\begin{equation}
\varepsilon \approx 0.00457
\quad\text{for } n=200{,}000.
\end{equation}
Thus, at $99\%$ confidence over the evaluated percentile grid, sampling $100{,}000$ coordinates confines each sampled percentile to a neighboring full-coordinate percentile band of about $0.65$ percentage points. Sampling $200{,}000$ coordinates tightens the band to about $0.46$ percentage points. These rank-space bounds complement the empirical calibration results: the observed sampled-coordinate boundary shape remains nearly identical to the full-coordinate boundary while substantially reducing percentile-computation cost.

\section{Full Boundary Stability Diagnostics}
\label{app:boundary-stability-full}

Table~\ref{tab:absolute-boundary-stability-full} reports the full set of boundary stability diagnostics used in \cref{sec:raw-boundary-validation}. The notation follows \cref{sec:protocol-boundary-calibration}: a raw boundary profile is a percentile function \(\widetilde B_s^{\rho}(p)\), where \(\rho\in\{\mathrm{abs},\mathrm{rel}\}\) and \(p\in\Lambda\). In this appendix, the subscript denotes the calibration units used to estimate the profile rather than the deployment stride. Thus, \(\widetilde B_n^{\rho}(p)\) is the raw profile estimated from the first \(n\) calibration units, and \(\widetilde B_{\star}^{\rho}(p)\) is the largest calibration-size reference profile for the same setting and profile type. We report diagnostics at \(\mathcal P=\{30,50,90\}\).

The SupNorm diagnostic measures relative drift from the reference profile:
\begin{equation}
\begin{aligned}
R_n^{\rho}(p)
&=
\frac{
|\widetilde B_n^{\rho}(p)
-\widetilde B_{\star}^{\rho}(p)|
}{
\widetilde B_{\star}^{\rho}(p)+\epsilon_{\mathrm{stab}}
},
\quad p\in\mathcal P,\\
\mathrm{SupNorm}^{\rho}(n)
&=
\max_{p\in\mathcal P} R_n^{\rho}(p).
\end{aligned}
\label{eq:boundary-stability-supnorm}
\end{equation}
This definition is consistent with the percentile-profile boundary in \cref{eq:raw-boundary}. The table reports the coordinate-wise terms \(R_n^{\rho}(p)\) under the SupNorm block, while the scalar SupNorm summary is their maximum over \(p\in\mathcal P\).

The remaining diagnostics test complementary sources of instability. Let \(\widetilde B_{n,-u}^{\rho}(p)\) be the profile obtained after removing calibration unit \(u\) from the first \(n\) units. The jackknife influence is
\begin{equation}
\mathrm{Jackknife}^{\rho}_n(p)
=
\max_{1\le u\le n}
\frac{
|\widetilde B_{n,-u}^{\rho}(p)
-\widetilde B_n^{\rho}(p)|
}{
\widetilde B_n^{\rho}(p)+\epsilon_{\mathrm{stab}}
}.
\label{eq:boundary-stability-jackknife}
\end{equation}
This diagnostic checks whether a single calibration unit dominates the estimated boundary coordinate.

For local variation along the calibration trajectory, fix a rolling-window size \(w\). Let \(\mathcal W_n=\{1,\ldots,n-w+1\}\) be the set of valid window starts, and let \(\widetilde B_{r:r+w-1}^{\rho}(p)\) be the profile estimated from the contiguous window starting at \(r\). The rolling standard-deviation diagnostic is
\begin{equation}
\mathrm{RollSD}^{\rho}_n(p)
=
\frac{
\operatorname{Std}_{r\in\mathcal W_n}
\left[
\widetilde B_{r:r+w-1}^{\rho}(p)
\right]
}{
\widetilde B_n^{\rho}(p)+\epsilon_{\mathrm{stab}}
}.
\label{eq:boundary-stability-rollsd}
\end{equation}
RollSD is therefore a normalized measure of short-range fluctuation in the calibration trajectory.

Finally, TailAdj measures how much the boundary changes when the most recent calibration units are incorporated. With the same tail length \(w\), define
\begin{equation}
\mathrm{TailAdj}^{\rho}_n(p)
=
\frac{
|\widetilde B_n^{\rho}(p)
-\widetilde B_{n-w}^{\rho}(p)|
}{
\widetilde B_n^{\rho}(p)+\epsilon_{\mathrm{stab}}
}.
\label{eq:boundary-stability-tailadj}
\end{equation}
A small TailAdj value indicates that the boundary profile has largely stopped moving at the end of the calibration run. Together, these diagnostics support the same conclusion as the main-text stability summary: short calibration runs already produce stable raw empirical boundary profiles.

A separate Qwen3-4B stride-aligned stability sweep covered training steps 1--16000. The same table includes this sweep for compactness. At stride $s=200$, the calibration grid contains 80 profiles; at stride $s=400$, it contains 40 profiles. 

\begin{table*}[t]
\centering
\caption{Full stability diagnostics of selected empirical boundary percentile profiles. Columns $p=30,50,90$ report the stability of the corresponding boundary-profile coordinates. For standard calibration rows, Size is the number of calibration steps; for Qwen3-4B stride rows, Size is the number of stride-aligned profiles. Values are relative changes, where $0.01$ corresponds to $1\%$; lower is better.}
\label{tab:absolute-boundary-stability-full}
\scriptsize
\setlength{\tabcolsep}{2.2pt}
\renewcommand{\arraystretch}{0.98}
\resizebox{\textwidth}{!}{%
\begin{tabular}{lllcccccccccccc}
\toprule
Setting & Size & Profile
& \multicolumn{3}{c}{SupNorm}
& \multicolumn{3}{c}{Jackknife}
& \multicolumn{3}{c}{RollSD}
& \multicolumn{3}{c}{TailAdj} \\
\cmidrule(lr){4-6}
\cmidrule(lr){7-9}
\cmidrule(lr){10-12}
\cmidrule(lr){13-15}
& & & $p=30$ & $p=50$ & $p=90$
  & $p=30$ & $p=50$ & $p=90$
  & $p=30$ & $p=50$ & $p=90$
  & $p=30$ & $p=50$ & $p=90$ \\
\midrule
\multirow{4}{*}{BERT-large}
& 100 & Abs. & 0.013 & 0.042 & 0.000 & 0.016 & 0.044 & 0.035 & 0.179 & 0.175 & 0.184 & 0.013 & 0.041 & 0.000 \\
& 200 & Abs. & 0.002 & 0.002 & 0.004 & 0.072 & 0.073 & 0.085 & 0.226 & 0.228 & 0.226 & 0.001 & 0.001 & 0.002 \\
& 300 & Abs. & 0.001 & 0.001 & 0.003 & 0.056 & 0.051 & 0.032 & 0.232 & 0.235 & 0.237 & 0.000 & 0.001 & 0.002 \\
& 400 & Abs. & 0.001 & 0.001 & 0.003 & 0.056 & 0.052 & 0.035 & 0.221 & 0.225 & 0.231 & 0.000 & 0.001 & 0.002 \\
\midrule
\multirow{4}{*}{Qwen3-4B}
& 100 & Abs. & 0.000 & 0.001 & 0.001 & 0.021 & 0.042 & 0.047 & 0.239 & 0.239 & 0.245 & 0.000 & 0.000 & 0.000 \\
& 200 & Abs. & 0.001 & 0.000 & 0.000 & 0.104 & 0.096 & 0.013 & 0.263 & 0.268 & 0.284 & 0.000 & 0.000 & 0.000 \\
& 300 & Abs. & 0.001 & 0.000 & 0.000 & 0.081 & 0.055 & 0.011 & 0.256 & 0.261 & 0.276 & 0.000 & 0.000 & 0.000 \\
& 400 & Abs. & 0.001 & 0.000 & 0.000 & 0.033 & 0.096 & 0.009 & 0.242 & 0.246 & 0.265 & 0.000 & 0.000 & 0.000 \\
\midrule
\multirow{4}{*}{ResNet-152}
& 100 & Abs. & 0.000 & 0.001 & 0.000 & 0.000 & 0.026 & 0.073 & 0.195 & 0.118 & 0.111 & 0.000 & 0.000 & 0.000 \\
& 200 & Abs. & 0.000 & 0.001 & 0.000 & 0.000 & 0.013 & 0.030 & 0.195 & 0.119 & 0.107 & 0.000 & 0.000 & 0.000 \\
& 300 & Abs. & 0.000 & 0.001 & 0.000 & 0.000 & 0.020 & 0.018 & 0.195 & 0.118 & 0.102 & 0.000 & 0.000 & 0.000 \\
& 400 & Abs. & 0.000 & 0.000 & 0.000 & 0.010 & 0.000 & 0.022 & 0.194 & 0.118 & 0.105 & 0.000 & 0.000 & 0.000 \\
\midrule
\multirow{4}{*}{Stable Diffusion 1.5}
& 100 & Abs. & 0.045 & 0.050 & 0.073 & 0.063 & 0.048 & 0.052 & 0.241 & 0.240 & 0.215 & 0.044 & 0.048 & 0.116 \\
& 200 & Abs. & 0.004 & 0.002 & 0.001 & 0.039 & 0.073 & 0.057 & 0.239 & 0.220 & 0.227 & 0.001 & 0.001 & 0.000 \\
& 300 & Abs. & 0.001 & 0.001 & 0.001 & 0.039 & 0.065 & 0.056 & 0.231 & 0.223 & 0.219 & 0.000 & 0.000 & 0.000 \\
& 400 & Abs. & 0.001 & 0.001 & 0.000 & 0.046 & 0.013 & 0.018 & 0.227 & 0.231 & 0.219 & 0.000 & 0.000 & 0.000 \\
\midrule
\multirow{6}{*}{Qwen3-4B ($s=200$)}
& 10 & Abs. & 0.000 & 0.000 & 0.000 & 0.059 & 0.168 & 0.339 & 0.259 & 0.252 & 0.247 & 0.000 & 0.000 & 0.000 \\
& 10 & Rel. & 0.000 & 0.000 & 0.000 & 0.802 & 0.830 & 0.714 & 0.263 & 0.267 & 0.251 & 0.000 & 0.000 & 0.000 \\
& 40 & Abs. & 0.000 & 0.000 & 0.000 & 0.230 & 0.126 & 0.037 & 0.272 & 0.286 & 0.307 & 0.000 & 0.000 & 0.000 \\
& 40 & Rel. & 0.000 & 0.000 & 0.000 & 0.415 & 0.553 & 0.547 & 0.215 & 0.186 & 0.165 & 0.000 & 0.000 & 0.000 \\
& 80 & Abs. & 0.000 & 0.000 & 0.000 & 0.230 & 0.126 & 0.037 & 0.256 & 0.270 & 0.293 & 0.000 & 0.000 & 0.000 \\
& 80 & Rel. & 0.000 & 0.000 & 0.000 & 0.415 & 0.553 & 0.547 & 0.200 & 0.168 & 0.152 & 0.000 & 0.000 & 0.000 \\
\midrule
\multirow{6}{*}{Qwen3-4B ($s=400$)}
& 5 & Abs. & 0.000 & 0.000 & 0.000 & 0.382 & 0.367 & 0.346 & 0.294 & 0.300 & 0.315 & 0.000 & 0.000 & 0.000 \\
& 5 & Rel. & 0.066 & 0.019 & 0.000 & 0.150 & 0.018 & 0.245 & 0.127 & 0.126 & 0.171 & 0.050 & 0.018 & 0.000 \\
& 20 & Abs. & 0.000 & 0.000 & 0.000 & 0.382 & 0.367 & 0.346 & 0.254 & 0.257 & 0.272 & 0.000 & 0.000 & 0.000 \\
& 20 & Rel. & 0.000 & 0.000 & 0.000 & 0.052 & 0.030 & 0.213 & 0.097 & 0.086 & 0.116 & 0.000 & 0.000 & 0.000 \\
& 40 & Abs. & 0.000 & 0.000 & 0.000 & 0.245 & 0.267 & 0.288 & 0.194 & 0.200 & 0.225 & 0.000 & 0.000 & 0.000 \\
& 40 & Rel. & 0.086 & 0.044 & 0.000 & 0.308 & 0.249 & 0.213 & 0.146 & 0.135 & 0.099 & 0.015 & 0.089 & 0.000 \\
\bottomrule
\end{tabular}%
}
\end{table*}

\section{Boundary Robustness Details}
\label{app:boundary-robustness}

This appendix reports the complete honest transfer results. All experiments use Qwen3-4B with honest training, SGD with learning rate $0.001$ and zero momentum, and stride $s=1$. Each configuration contains 1,100 ordered intervals. Intervals 1--100 are used only for calibration, while intervals 101--1100 form a disjoint 1,000-interval judgment trace.  The provider uses an RTX PRO 4000 and replay is performed on an NVIDIA L4. For the hardware transfer experiment, replay is independently performed on an L4, RTX 4090, and RTX PRO 4500. In all tables below, rows denote the source configuration used to calibrate the boundary and columns denote the honest target judgment trace. Each entry reports the number of rejected intervals out of 1,000.

\subsection{LoRA Layer-Range Transfer}

We vary only the range of transformer layers adapted by LoRA: layers 20--35, 24--35, and 28--35. The dataset, maximum input length, LoRA module types, optimizer, sample order, and provider/verifier hardware remain fixed. 

\begin{table}[htbp]
    \centering
    \caption{Honest rejection counts for LoRA layer-range transfer.
    Each entry is rejected intervals out of 1,000.}
    \label{tab:boundary-transfer-layers}
    \begin{tabular}{lccc}
        \toprule
        Boundary source
        & L20--35
        & L24--35
        & L28--35 \\
        \midrule
        L20--35 & 0 & 1 & 0 \\
        L24--35 & 0 & 1 & 0 \\
        L28--35 & 0 & 1 & 0 \\
        \bottomrule
    \end{tabular}
\end{table}

\subsection{Dataset Transfer}

We next vary the fine-tuning dataset between NuminaMath-CoT and Alpaca while keeping the 512-token limit, LoRA layers 20--35, optimizer, and hardware roles fixed.

\begin{table}[htbp]
    \centering
    \caption{Honest rejection counts for dataset transfer.
    Each entry is rejected intervals out of 1,000.}
    \label{tab:boundary-transfer-dataset}
    \begin{tabular}{lcc}
        \toprule
        Boundary source
        & NuminaMath
        & Alpaca \\
        \midrule
        NuminaMath & 0 & 11 \\
        Alpaca     & 0 & 0  \\
        \bottomrule
    \end{tabular}
\end{table}

The transfer is asymmetric: the NuminaMath-CoT boundary rejects 11 Alpaca intervals, giving an FRR of $1.1\%$, while the Alpaca boundary produces no observed rejection on either target. 

\subsection{Input-Length Transfer}

We vary the maximum input length among 256, 512, and 1024 tokens while holding the dataset, LoRA configuration, optimizer, sample order, and hardware roles fixed.

\begin{table}[htbp]
    \centering
    \caption{Honest rejection counts for maximum-input-length transfer.
    Each entry is rejected intervals out of 1,000.}
    \label{tab:boundary-transfer-tokens}
    \begin{tabular}{lccc}
        \toprule
        Boundary source
        & 256
        & 512
        & 1024 \\
        \midrule
        256  & 1 & 0 & 0 \\
        512  & 1 & 0 & 0 \\
        1024 & 1 & 0 & 0 \\
        \bottomrule
    \end{tabular}
\end{table}

Every boundary rejects the same 256-token judgment trace once, including the 256-token diagonal boundary. All 512- and 1024-token judgment traces have zero observed false rejections. 

\subsection{Verifier-GPU Transfer}

Finally, we fix the provider checkpoints and the complete training configuration and vary only the verifier GPU. Each verifier independently calibrates a boundary from intervals 1--100 and evaluates intervals 101--1100.

\begin{table}[htbp]
    \centering
    \caption{Honest rejection counts for verifier-GPU boundary transfer.
    Each entry is rejected intervals out of 1,000.}
    \label{tab:boundary-transfer-hardware}
    \begin{tabular}{lccc}
        \toprule
        Boundary source
        & L4
        & RTX 4090
        & PRO 4500 \\
        \midrule
        L4       & 0 & 0 & 0 \\
        RTX 4090 & 0 & 0 & 0 \\
        PRO 4500 & 0 & 0 & 0 \\
        \bottomrule
    \end{tabular}
\end{table}

\section{Baseline Verifier-Signal Details}
\label{app:baseline-details}

This appendix records the score definitions used for the signal-level baselines
in \cref{sec:baseline-comparison}. These instantiations are intended to compare
verification evidence under the same target-perturbation attack, not to
reproduce the full deployment protocols of the original systems.

\subsection{PoL-L2 Endpoint Distance}

PoL-L2 uses the same honest calibration units as
\cref{sec:protocol-boundary-calibration}, but replaces the gradient-percentile
boundary with a scalar endpoint-distance threshold. For a replay unit
\(\tau=[a,b)\), let \(W_b^{\mathrm{ref}}\) be the checkpoint obtained by honest
replay and let \(W_b^{\mathrm{open}}(\lambda)\) be the endpoint after adding the
target direction at scale \(\lambda\). We set
\begin{equation}
\gamma_{\mathrm{PoL}}
=
\max_{\tau\in\mathcal{C}_s}
\left\|
W_{b,\tau}^{\mathrm{honest},1}
-
W_{b,\tau}^{\mathrm{honest},2}
\right\|_2,
\end{equation}
using honest calibration endpoints for the same replay setting. The PoL-L2
predicate accepts when
\begin{equation}
\mathcal{C}_{\mathrm{PoL}}(\lambda)
=
\mathbf{1}\!\left[
\left\|
W_b^{\mathrm{open}}(\lambda)-W_b^{\mathrm{ref}}
\right\|_2
\leq
\gamma_{\mathrm{PoL}}
\right].
\end{equation}

\subsection{PoTD Segment Score}

PoTD checks whether a claimed training segment exhibits the expected
training-data memorization signal. Let \(\Pi_i\) be the claimed data segment
between checkpoints \(W_{i-1}\) and \(W_i\), and let \(D_{\mathrm{val}}\) be a
held-out validation set. For a sample \(d\) and checkpoint \(W\), define the
relative memorization score
\begin{equation}
M(d,W)
=
\frac{1}{|D_{\mathrm{val}}|}
\sum_{d'\in D_{\mathrm{val}}}
\mathcal{L}(d',W)
-
\mathcal{L}(d,W).
\end{equation}
The segment-level change is
\begin{equation}
\Delta M(d,i)
=
M(d,W_i)-M(d,W_{i-1}).
\end{equation}
For \(p\in\{0.1,0.2\}\), let \(q_i^{\mathrm{val}}(p)\) be the \(p\)-quantile of
\(\{\Delta M(d,i):d\in D_{\mathrm{val}}\}\). The fraction-below-quantile score
for a set \(A\) is
\begin{equation}
\mathrm{FBQ}(A,p,i)
=
\frac{1}{|A|}
\sum_{d\in A}
\mathbf{1}\!\left[
\Delta M(d,i)
\leq
q_i^{\mathrm{val}}(p)
\right].
\end{equation}
We instantiate the PoTD segment statistic as
\begin{equation}
S_{\mathrm{PoTD},i}
=
\max_{p\in\{0.1,0.2\}}
\frac{
\mathrm{FBQ}(\Pi_i,p,i)
}{
\mathrm{FBQ}(D_{\mathrm{val}},p,i)+\epsilon
}.
\end{equation}
The baseline uses \(10\)-step segments and enforces the check at segment
boundaries. The threshold \(\gamma_{\mathrm{PoTD}}\) is calibrated from the clean
reference transcript, and the predicate accepts when
\begin{equation}
\mathcal{C}_{\mathrm{PoTD}}(\lambda)
=
\mathbf{1}\!\left[
\max_i S_{\mathrm{PoTD},i}(\lambda)
\leq
\gamma_{\mathrm{PoTD}}
\right].
\end{equation}

\subsection{RTTD Replicated-Execution Score}

RTTD compares the primary server's audited sub-run with clean replicated
executions. In our instantiation, server \(0\) is the primary attacked server,
and servers \(1\)--\(6\) are clean replicas initialized from the same checkpoint.
Each audited sub-run has length \(10\). After the sub-run, each server returns an
endpoint checkpoint \(W_{\mathrm{end}}^{(r)}\).

We compute a Zest-style behavior signature \(\phi(W)\) by masking token segments
of reference samples, recording the supervised-label logit response, and fitting
a local linear surrogate. The distance between two endpoints is cosine distance
in this signature space:
\begin{equation}
D_{\mathrm{Zest}}(W_a,W_b)
=
1-
\frac{
\langle \phi(W_a),\phi(W_b)\rangle
}{
\|\phi(W_a)\|_2\|\phi(W_b)\|_2
}.
\end{equation}
The clean cluster is estimated from the replica--replica distances, and the
primary-to-replica distances are compared with this cluster using a two-sample
Kolmogorov--Smirnov test. Let \(p_{\mathrm{Zest/KS}}(\lambda)\) denote the
largest resulting p-value over the primary comparison windows. We use
\(\alpha_{\mathrm{KS}}=0.01\) as the rejection significance level, so the RTTD
predicate is
\begin{equation}
\mathcal{C}_{\mathrm{RTTD}}(\lambda)
=
\mathbf{1}\!\left[
p_{\mathrm{Zest/KS}}(\lambda)
\geq
\alpha_{\mathrm{KS}}
\right].
\end{equation}

\end{document}